\documentclass[aps,prd,preprint,groupedaddress]{revtex4-1}

\usepackage{graphicx}
\usepackage{epstopdf}
\usepackage{amsmath}

\usepackage{hyperref}

\usepackage{pstricks} 

\newcommand{\mbf}[1]{\mathbf{#1}}

\begin{document}

\preprint{SLAC--PUB--13840 ~~~
JLAB--PHY--10--1128}

\title{Nonperturbative QCD Coupling and its $\beta$ Function  from Light-Front Holography}

\author{Stanley J. Brodsky}
\affiliation{SLAC National Accelerator Laboratory, Stanford University,
Stanford, California 94309, USA}
\affiliation{CP$^3$-Origins, University of Southern Denmark, Odense, 5230 M,  Denmark}

\author{Guy F.  de T\'eramond}
\affiliation{Universidad de Costa Rica, San Jos\'e, Costa Rica}

\author{Alexandre Deur}
\affiliation{Thomas Jefferson National Accelerator Facility, Newport News, VA 23606}

\date{\today}

\begin{abstract}

The light-front holographic mapping of classical gravity in anti-de Sitter space, modified by a positive-sign 
dilaton background,  leads to a nonperturbative effective coupling 
$\alpha_s^{AdS}(Q^2)$. It agrees with hadron physics data
extracted from different observables, such as the effective charge defined by the Bjorken sum rule, as well as with the predictions of models with built-in confinement  and 
lattice simulations. It also displays a transition from  perturbative to nonperturbative conformal regimes
at a momentum scale $ \sim 1$ GeV.  The resulting   $\beta$ function  appears to capture the essential characteristics of the full 
 $\beta$ function of QCD, thus giving further support to the application of the gauge/gravity duality to the 
 confining dynamics of strongly coupled QCD.  Commensurate scale relations relate observables to each other without scheme or scale  ambiguity.  In this paper we extrapolate these relations 
to the nonperturbative domain,  thus extending the range of predictions based on $\alpha_s^{AdS}(Q^2)$.

\end{abstract}



%\keywords{}

\maketitle

\section{Introduction \label{intro}}

The concept of a running coupling $\alpha_s(Q^2)$  in QCD is usually restricted to the perturbative domain.  However, as in QED, it is useful to define the coupling as an analytic function valid over the full spacelike and timelike domains.
The study of the non-Abelian QCD coupling at small momentum transfer is a complex problem because of  gluonic self-coupling and color confinement.  Its behavior in the nonperturbative infrared (IR) regime has  been the subject of intensive study using Dyson-Schwinger equations  and Euclidean numerical lattice computation,~\cite{Binosi:2009qm} since it is a quantity  of fundamental importance.
We will  show that the light-front (LF) holographic mapping of classical gravity in anti-de Sitter (AdS) space, modified by a positive-sign 
dilaton background $\exp{(+ \kappa^2 z^2)}$, leads to a nonperturbative effective coupling 
$\alpha_s^{AdS}(Q^2)$ which is in agreement with hadron physics data
extracted from different observables, as well as with  the predictions of models with built-in confinement  and 
lattice simulations.

The AdS/CFT correspondence~\cite{Maldacena:1997re} between a gravity or string theory on a higher dimensional AdS space-time and conformal gauge field theories  in physical space-time has
brought  a new set of tools for studying the dynamics of strongly coupled quantum field theories, and it has led
to new analytical insights into the confining dynamics of QCD.  
The AdS/CFT duality provides a gravity
description in a ($d+1$)-dimensional AdS
spacetime in terms of a flat
$d$-dimensional conformally-invariant quantum field theory defined at the AdS 
asymptotic boundary.~\cite{Gubser:1998bc}
Thus, in principle, one can compute physical observables in a strongly coupled gauge theory  in terms of a classical gravity theory. 

Since
the quantum field theory dual to AdS$_5$ space in the original correspondence~\cite{Maldacena:1997re} is conformal,
the strong coupling of the dual gauge theory is constant, and its $\beta$ function is zero. Thus, one must consider a deformed AdS space in order to have a running coupling $\alpha_s^{AdS}(Q^2)$ for the gauge theory side of the correspondence.   We  assume a  positive-sign confining dilaton background to modify AdS space, a model
that gives a very good account of meson and baryon spectroscopy and form factors.
We use LF holography \cite{deTeramond:2008ht, Brodsky:2006uqa, Brodsky:2008pf, deTeramond:2009xk, deTeramond:2010we}   to  map the amplitudes corresponding to
hadrons  propagating in AdS space to the frame-independent  light-front wave functions (LFWFs)
of hadrons in physical $3+1$ space. This analysis utilizes recent developments in LF QCD, which have been inspired by the AdS/CFT correspondence.~\cite{Maldacena:1997re} The resulting LFWFs provide a fundamental description of the structure and internal dynamics of hadronic states in terms of their constituent quarks and gluons.

The definition of the running coupling in perturbative quantum field theory is scheme-dependent.  As discussed by Grunberg,~\cite{Grunberg}  an effective coupling or charge can be defined directly from physical observables. 
Effective charges defined from
different observables can be related  to each other in the leading-twist domain using commensurate scale relations
 (CSR).~\cite{CSR}  A more challenging problem is to relate such observables and schemes over the full domain of momenta. An important part of this paper will be the application and test of commensurate scale relations and  their tentative extension to the nonperturbative domain. Another important application is related to the  potential between infinitely heavy quarks,
 which can be defined analytically in momentum transfer
 space as the product  of the running coupling times the Born gluon propagator: $V(q)  = - 4 \pi C_F {\alpha_V(q) / q^2}$.   This effective charge defines a renormalization scheme -- the $\alpha_V$ scheme of Appelquist, Dine,  and Muzinich.~\cite{Appelquist:1977tw}  
In fact, the holographic coupling $\alpha_s^{AdS}(Q^2)$ can be considered to be the nonperturbative extension of the 
$\alpha_V$ effective charge defined in Ref. \cite{Appelquist:1977tw}.

We shall also make extensive use of the $g_1$ scheme, where the strong coupling $\alpha_{g_1}(Q^2)$ is determined from
the Bjorken sum rule.~\cite{Bjorken SR}  The coupling $\alpha_{g_1}(Q^2)$ has the advantage that it is the best-measured effective charge, and it can be used to extrapolate the definition of the effective coupling to large distances.~\cite{Deur:2009hu} It has  been measured at intermediate energies, and it is therefore
particularly useful to study  the transition from short distances, where partons are the relevant degrees of freedom,  to large distances, where the hadronic degrees of freedom are present.~\cite{FFexample}

This paper is organized as follows: after briefly reviewing in Sec. \ref{lfholography} the light-front quantization approach to the
gauge/gravity correspondence, we identify a nonperturbative running coupling in Sec. \ref{alphaAdS} from the fifth-dimensional action  of gauge fields propagating in AdS$_5$ space  modified by a positive-sign 
dilaton background $\exp{(+ \kappa^2 z^2)}$.  In Sec. \ref{alphatest}, we compare the results for the coupling
$\alpha^{AdS}_s$ obtained in
Sec.   \ref{alphaAdS}  with the effective QCD couplings  extracted from
different observables and lattice results.  The nonperturbative results are extended  to large $Q^2$ 
by matching the holographic results to the perturbative results in the transition region.
In Sec. \ref{betaAdS}, we discuss the holographic  results for the $\beta$ function in the nonperturbative domain and compare the predictions with lattice and experimental results.  In Sec. \ref{csr}, we discuss the use of CSR to relate  different 
effective charges. A discussion of experimental results, schemes and data normalization is given in Sec. \ref{exp}. 
The CSR discussion is extended in Sec.  \ref{alphar} to configuration space. Some final remarks
are given in the conclusions in Sec. \ref{conclusions}. A check on the validity of CSR is carried out in the Appendix
where the results for  the $g_1$, $V$, and $\overline{MS}$ schemes are confronted in the perturbative domain.

\section{Light-Front Holography  and QCD\label{lfholography}}

The basic principle underlying the AdS/CFT approach to conformal gauge theories is the isomorphism of the group of
Poincar\'e  and conformal transformations $SO(4,2)$ to the group of isometries of AdS$_5$ space,
the group of transformations that leave the AdS metric
\begin{equation} \label{eq:AdSz}
ds^2 = \frac{R^2}{z^2} \left(\eta_{\mu \nu} dx^\mu dx^\nu - dz^2\right),
\end{equation}
invariant ($R$ the AdS radius).  Since the metric  (\ref{eq:AdSz})
is invariant under a dilatation of all coordinates $x^\mu \to \lambda x^\mu$, $z \to \lambda z$, the variable $z$ acts like a scaling variable in Minkowski space: different values of $z$ correspond to different energy scales at which the hadron is examined. 

In order to describe a confining theory,
 the conformal invariance of AdS$_5$ must be broken. A simple way to impose confinement and  discrete
normalizable modes is to truncate the regime where the string modes can propagate by introducing an IR cutoff at a finite value   $z_0 \sim 1/\Lambda_{\rm QCD}$. Thus, the ``hard-wall'' at $z_0$ breaks conformal invariance and allows the introduction of the QCD scale  and a spectrum of particle states.~\cite{Polchinski:2001tt} In this simplified  approach
the propagation
of hadronic modes in a fixed effective gravitational background encodes the salient properties of the QCD dual theory, such
as the ultraviolet (UV) conformal limit at the AdS boundary at $z \to 0$, as well as modifications of the background geometry in the large $z$ infrared region which are dual to confining gauge theories.  As first shown by Polchinski
and Strassler,~\cite{Polchinski:2001tt} the AdS/CFT duality, modified
to incorporate a mass scale, 
provides a derivation of dimensional counting
rules~\cite{Brodsky:1973kr} for the leading 
power-law falloff of hard scattering beyond the perturbative regime.
The modified theory generates the hard behavior expected from QCD, instead of the soft
behavior characteristic of strings. 

The conformal metric of AdS space can be modified  within the AdS/QCD framework  to simulate  confinement 
forces.~\cite{Karch:2006pv} The
introduction of a dilaton profile in the AdS action can be considered  equivalent to modifying  the AdS metric (\ref{eq:AdSz})
by introducing 
an additional warp factor  $\exp{\left(\pm \kappa^2 z^2\right)}$~\cite{Andreev:2006ct}
 \begin{equation}
 ds^2 = 
 \frac{R^2}{z^2} \, {e^{\pm \kappa^2 z^2}}  \left(\eta_{\mu \nu}  dx^\mu dx^\nu \! -  dz^2\right).
 \end{equation}
A dilaton profile $\exp{\left(\pm \kappa^2 z^2\right)}$ of either sign also leads to a
two-dimensional oscillator potential $U(\zeta) \sim \kappa^4 \zeta^2$ in the
relativistic LF eigenvalue equation of Ref.~\cite{deTeramond:2008ht}, which in turn 
reproduces the observed linear Regge trajectories in a Chew-Frautschi plot.  Glazek and Schaden~\cite{Glazek:1987ic} have shown that in QCD a  harmonic oscillator confining
potential naturally arises as an effective potential between heavy quark states when
higher gluonic Fock states are stochastically eliminated. 

The modified metric induced by the dilaton can be interpreted in AdS space as a gravitational potential 
for an object of mass $m$  in the fifth dimension:
$V(z) = mc^2 \sqrt{g_{00}} = mc^2 R \, e^{\pm \kappa^2 z^2/2}/z$.
In the case of the negative solution the potential decreases monotonically, and thus an object in AdS will fall to infinitely large 
values of $z$.  For the positive solution, the potential is nonmonotonic and has an absolute minimum at $z_0 = 1/\kappa$.  
Furthermore, for large values of $z$ the gravitational potential increases exponentially, thus confining any object  to distances $\langle z \rangle \sim 1/\kappa$.~\cite {deTeramond:2009xk}  We thus use the positive-sign dilaton solution opposite to that of Ref. \cite{Karch:2006pv}. This additional warp factor leads to a well-defined scale-dependent effective coupling.
Introducing a positive-sign dilaton background is also relevant for describing chiral symmetry breaking in the soft-wall model, \cite{Gherghetta:2009ac}  since  the expectation value of the scalar field associated with the quark mass and condensate does not blow up in the far infrared region of 
AdS,~\cite{Zuo:2009dz}  in contrast with the original model.~\cite{Karch:2006pv}

The soft-wall model of Ref. \cite{Karch:2006pv} also uses the AdS/QCD framework~\cite{Erlich:2005qh, DaRold:2005zs}, 
where bulk fields are introduced to match the 
$SU(2)_L \times SU(2)_R$ chiral symmetry of QCD and its spontaneous breaking, but without  an explicit connection to the internal constituent structure of hadrons.~\cite{Brodsky:2003px}  Instead, axial and vector currents become the
primary entities as in an effective chiral theory.  In this ``bottom-up" model
only a limited number of operators  are introduced, and consequently, only a limited
number of fields are required to construct  phenomenologically viable five-dimensional gravity duals.

Light-front  holography provides a remarkable
connection between the equations of motion in AdS space and
the Hamiltonian formulation of QCD in physical spacetime quantized
on the light front  at fixed LF time  $\tau = x^+ = x^0 + x^3$, the time marked by the
front of a light wave.~\cite{Dirac:1949cp}   This correspondence provides a direct connection between the hadronic amplitudes $\Phi(z)$  in AdS space  with  LFWFs $\phi(\zeta)$ describing the quark and gluon constituent structure of hadrons in physical space-time.
The mapping between the LF invariant variable $\zeta$ and the fifth-dimension AdS coordinate $z$ was originally obtained
by matching the expression for electromagnetic (EM) current matrix
elements in AdS space with the corresponding expression for the
current matrix element, using LF  theory in physical spacetime.~\cite{Brodsky:2006uqa}   It has also been shown that one
obtains the identical holographic mapping using the matrix elements
of the energy-momentum tensor,~\cite{Brodsky:2008pf} thus  verifying  the  consistency of the holographic
mapping from AdS to physical observables defined on the light front.  
LF holography thus provides a direct correspondence between an effective gravity theory defined in a fifth-dimensional warped space and a physical description of hadrons in  $3+1$ spacetime.

Light-front quantization is the ideal framework for  describing the
structure of hadrons in terms of their quark and gluon degrees of
freedom. LFWFs  play the same role in
hadron physics that Schr\"odinger wave functions play in atomic physics.~\cite{Brodsky:1997de} 
The simple structure of the LF vacuum provides an unambiguous
definition of the partonic content of a hadron in QCD. 
A physical hadron in four-dimensional Minkowski space has four-momentum $P_\mu$ and invariant
hadronic mass states, $P_\mu P^\mu = \mathcal{M}^2$, determined by the 
Lorentz-invariant Hamiltonian equation for the relativistic bound-state system
\begin{equation} \label{eq:LFH}
P_\mu P^\mu \vert  \psi(P) \rangle = \left( P^- P^+ \!  - 
 \mbf{P}_\perp^2\right) \vert \psi(P) \rangle =
 \mathcal{M}^2 \vert  \psi(P) \rangle.
 \end{equation}
 The hadron  four-momentum generator  is $P = (P^+, P^-, \mbf{P}_{\! \perp})$, $P^\pm = P^0 \pm P^3$, and
 the hadronic state $\vert\psi\rangle$ is an expansion in multiparticle Fock eigenstates
$\vert n \rangle$ of the free light-front  Hamiltonian: 
$\vert \psi \rangle = \sum_n \psi_n \vert n\rangle$. 
The internal partonic coordinates of the hadron  are the momentum fractions $x_i = k^+_i/P^+$ and the transverse momenta $\mbf{k}_{\perp i}$, $i = 1, 2,\dots, n$,
where $n$ is the number of partons in a given Fock state. 
Momentum conservation requires 
$\sum_{i=1}^n x_i = 1$ and $\sum_{i=1}^n \mbf{k}_{\perp i}= 0$.  It is useful to employ a mixed 
representation~\cite{Soper:1976jc} in terms of 
 $n-1$ independent momentum fraction variables $x_j$ and position coordinates $\mbf{b}_{\perp j}$, $j = 1, 2, \dots, n-1$,
so that $\sum_{i=1}^n \mbf{b}_{\perp i}= 0$. The relative transverse variables $\mbf{b}_{\perp i}$ are Fourier conjugates
of the momentum variables $\mbf{k}_{\perp i}$. 

In AdS space the physical  states are
represented by normalizable modes
$\Phi_P(x^\mu,z) = e^{-iP \cdot x} \Phi(z)$,
with plane waves along the Poincar\'e coordinates and a profile function $\Phi(z)$ 
along the holographic coordinate $z$. Each  LF hadronic state $\vert \psi(P) \rangle$ is dual to a normalizable string mode $\Phi_P(x^\mu,z)$.  The hadronic mass  $\mathcal{M}^2$ is found by solving the eigenvalue problem for the corresponding wave equation in AdS space, which, as we discuss below, is equivalent to the semiclassical approximation 
to the light-front bound-state Hamiltonian equation of motion in QCD. 
One can indeed systematically reduce  the LF  Hamiltonian eigenvalue Eq.  (\ref{eq:LFH}) to an effective relativistic wave equation~\cite{deTeramond:2008ht}  by observing that each $n$-particle Fock state has an essential dependence on the invariant mass of the system $M_n^2  = \left( \sum_{a=1}^n k_a^\mu\right)^2$ and thus, to a first approximation, LF dynamics depend only on $M_n^2$. In  impact space the relevant variable is the boost invariant transverse variable $\zeta$
 which measures the
separation of the quark and gluonic constituents within the hadron
at the same LF time and which also allows one to separate the dynamics
of quark and gluon binding from the kinematics of the constituent
internal angular momentum. 
In the case of two constituents, $\zeta = \sqrt{x(1-x)} \vert \mbf{b}_\perp \vert$  where 
$x = k^+/P^+ = (k^0 + k^3)/ (P^0 + P^3)$ is the LF fraction. The result is the single-variable light-front relativistic
Schr\"odinger equation~\cite{deTeramond:2008ht}
\begin{equation} \label{eq:SLFWE}
\left(-\frac{d^2}{d\zeta^2}
- \frac{1 - 4L^2}{4\zeta^2} + U(\zeta) \right)
\phi(\zeta) = \mathcal{M}^2 \phi(\zeta),
\end{equation}
where $U(\zeta)$ is the effective potential, and 
$L$ is the relative orbital angular momentum as defined in the LF formalism.
The set of eigenvalues  $\mathcal{M}^2$ gives the hadronic spectrum of the color-singlet states, and the corresponding eigenmodes  $\phi(\zeta)$ represent the LFWFs,  which describe the dynamics of the constituents of the hadron.   
This first approximation to relativistic QCD bound-state systems is 
equivalent to the equations of motion, which describe the propagation of spin-$J$ modes in a fixed  gravitational background asymptotic to AdS space.~\cite{deTeramond:2008ht}   By using the correspondence between $\zeta$ in the LF theory and $z$ in AdS space, one can identify the terms in the dual gravity AdS equations, which correspond to the kinetic energy terms of  the partons inside a hadron and the interaction terms that build confinement.~\cite{deTeramond:2008ht}  The identification of orbital angular momentum of the constituents in the light-front description is also a key element in our description of the internal structure of hadrons using holographic principles.

As we will discuss, the conformal AdS$_5$  metric (\ref{eq:AdSz}) can be deformed by a warp
factor $\exp{\left(+ \kappa^2 z^2\right)}$.  In the case of a two-parton relativistic bound state, the resulting effective potential in the LF equation of motion is 
$U(\zeta) = \kappa^4 \zeta^2 + 2 \kappa^2(L+S-1)$.~\cite{deTeramond:2009xk} There is only one parameter, the mass scale $\kappa \sim 1/2$ GeV, which enters the effective confining harmonic oscillator potential.  Here $S=0,1$ is the spin of the
quark-antiquark  system, $L$ is their relative orbital angular momentum, and $\zeta$  is the Lorentz-invariant coordinate defined above, which measures 
the distance between the quark and antiquark; it is analogous to the radial coordinate $r$ in the Schr\"odinger equation.  The resulting mesonic spectrum has the phenomenologically successful Regge form $\mathcal{M}^2 = 4 \kappa^2 (n+L +S/2),$ with equal slopes in the orbital angular momentum and the radial quantum number $n$. 
The pion with $n=L=S=0$ is massless for zero quark mass, consistent with chiral symmetry.

\section{Nonperturbative  QCD Coupling from Light-Front Holography \label{alphaAdS}}

We will show in this section how the LF holographic mapping of effective classical gravity in AdS space, modified by a positive-sign dilaton background, can  be used to identify an analytically simple  color-confining
nonperturbative effective coupling $\alpha_s^{AdS}(Q^2)$ as a function of the spacelike momentum transfer $Q^2 = - q^2$.   As we shall show, this coupling incorporates  confinement
and agrees well with effective charge observables and lattice simulations.  
It also exhibits an infrared fixed point at small $Q^2$ and asymptotic freedom at large $Q^2$. However, the falloff   of 
$\alpha_s^{AdS}(Q^2)$  at large $Q^2$ is exponential: $\alpha_s^{AdS}(Q^2) \sim e^{-Q^2 /  \kappa^2}$, rather than the perturbative QCD (pQCD) logarithmic falloff.   We shall show in later sections that a phenomenological extended coupling can be defined which implements the pQCD behavior.

As will be explained in Sec. \ref{betaAdS}, the  $\beta$ function derived from light-front holography becomes significantly  negative in the nonperturbative regime $Q^2 \sim \kappa^2$, where it reaches a minimum, signaling the transition region from the IR conformal region, characterized by hadronic degrees of freedom,  to a pQCD conformal UV  regime where the relevant degrees of freedom are the quark and gluon constituents.  The  $\beta$ function is always negative; it vanishes at large $Q^2$ consistent with asymptotic freedom, and it vanishes at small $Q^2$ consistent with an infrared fixed point.~\cite{Cornwall:1981zr, Brodsky:2008be}

Let us consider a five-dimensional gauge field $F$ propagating in AdS$_5$ space in the presence of a dilaton background 
$\varphi(z)$ which introduces the energy scale $\kappa$ in the five-dimensional action.
At quadratic order in the field strength the action is
\begin{equation}
S =  - {1\over 4}\int \! d^5x \, \sqrt{g} \, e^{\varphi(z)}  {1\over g^2_5} \, F^2,
\label{eq:action}
\end{equation}
where the metric determinant of AdS$_5$ is $\sqrt g = ( {R/z})^5$,  $\varphi=  \kappa^2 z^2$, and the square of the coupling $g_5$ has dimensions of length.   On general grounds we would expect that the value of the five-dimensional coupling $g_5^2$ in units $R=1$ is determined by a geometrical
factor scaled by $1/N_C$. We  can identify the prefactor 
\begin{equation} \label{eq:flow}
g^{-2}_5(z) =  e^{\varphi(z)}  g^{-2}_5 ,
\end{equation}
in the  AdS  action (\ref{eq:action})  as the effective coupling of the theory at the length scale $z$. 
The coupling $g_5(z)$ then incorporates the nonconformal dynamics of confinement. The five-dimensional coupling $g_5(z)$
is mapped,  modulo a  constant, into the Yang-Mills (YM) coupling $g_{\rm YM}$ of the confining theory in physical space-time using light-front holography. One  identifies $z$ with the invariant impact separation variable $\zeta$ which appears in the LF Hamiltonian:
$g_5(z) \to g_{\rm YM}(\zeta)$. Thus 
\begin{equation}  \label{eq:gYM}
\alpha_s^{AdS}(\zeta) = g_{\rm YM}^2(\zeta)/4 \pi \propto  e^{-\kappa^2 \zeta^2} . 
\end{equation}

In contrast with the three-dimensional radial coordinates of the nonrelativistic Schr\"odinger theory, the natural light-front
variables are the two-dimensional cylindrical coordinates $(\zeta, \phi)$ and the
light-cone fraction $x$. The physical coupling measured at the scale $Q$ is the two-dimensional Fourier transform
of the  LF transverse coupling $\alpha_s^{AdS}(\zeta)$  (\ref{eq:gYM}). Integration over the azimuthal angle
 $\phi$ gives the Bessel transform
 \begin{equation} \label{eq:2dimFT}
\alpha_s^{AdS}(Q^2) \sim \int^\infty_0 \! \zeta d\zeta \,  J_0(\zeta Q) \, \alpha_s^{AdS}(\zeta),
\end{equation}
in the $q^+ = 0$ light-front frame where $Q^2 = -q^2 = - \mbf{q}_\perp^2 > 0$ is the square of the spacelike 
four-momentum transferred to the 
hadronic bound state.   Using this ansatz we then have from  Eq.  (\ref{eq:2dimFT})
\begin{equation}
\label{eq:alphaAdS}
\alpha_s^{AdS}(Q^2) = \alpha_s^{AdS}(0) \, e^{- Q^2 /4 \kappa^2}.
\end{equation}
In contrast, the negative dilaton solution $\varphi=  -\kappa^2 z^2$ leads to an integral that diverges at large $\zeta$. 
The essential assumption of this paper is the identification of $\alpha_s^{AdS}(Q^2)$ with the physical QCD running coupling
in its nonperturbative domain.

The flow Eq.  (\ref{eq:flow}) from the scale-dependent measure for the gauge fields can be understood as a consequence of field-strength renormalization.
In physical QCD we can rescale the non-Abelian gluon field  $A^\mu \to \lambda A^\mu$  and field strength
$G^{\mu \nu}  \to \lambda G^{\mu \nu}$  in the QCD Lagrangian density  $\mathcal{L}_{\rm QCD}$ by a compensating rescaling of the coupling strength $g \to \lambda^{-1} g.$  The renormalization of the coupling $g _{phys} = Z^{1/2}_3  g_0,$   where $g_0$ is the bare coupling in the Lagrangian in the UV-regulated theory,  is thus  equivalent to the renormalization of the vector potential and field strength: $A^\mu_{ren} =  Z_3^{-1/2} A^\mu_0$, $G^{\mu \nu}_{ren} =  Z_3^{-1/2} G^{\mu \nu}_0$   with a rescaled Lagrangian density 
${\cal L}_{\rm QCD}^{ren}  = Z_3^{-1}  { \cal L}_{\rm QCD}^0  = (g_{phys}/g_0)^{-2}  \mathcal{L}_0$.  
 In lattice gauge theory,  the lattice spacing $a$ serves as the UV regulator, and the renormalized QCD coupling is determined  from the normalization of the gluon field strength as it appears  in the gluon propagator. The inverse of the lattice size $L$ sets the mass scale of the resulting running coupling.
As in lattice gauge theory, color confinement in AdS/QCD reflects nonpertubative dynamics at large distances. The QCD couplings defined from lattice gauge theory and the soft-wall holographic model are thus similar in concept, and both schemes are expected to have similar properties in the nonperturbative domain, up to a rescaling of their respective momentum scales.

The gauge/gravity correspondence has also been used to study the running coupling of the dual field theory. 
One can modify the dynamics of the dilaton in the AdS space to simulate the QCD $\beta$ function in the UV 
domain.~\cite{Csaki:2006ji, Hirn:2005vk, Gursoy:2007cb, Zeng:2008sx, Galow:2009kw, Alanen:2009xs, Jarvinen:2009fe}  For example, a 
$\beta$-function ansatz of the boundary field theory is used as input in Refs.~\cite{Gursoy:2007cb, Zeng:2008sx, Galow:2009kw, Alanen:2009xs, Jarvinen:2009fe} to modify the AdS metrics assuming the correspondence  between the AdS variable $z$ and the energy scale $E$ of the conformal field theory,  $E \sim 1/z$, as discussed in Ref.~\cite{Peet:1998wn}. In our paper,   the effective QCD coupling is identified by using the precise mapping from $z$ in AdS space to the transverse impact variable $\zeta$ in LF QCD.

\section{Comparison of  the Holographic Coupling with Other Effective Charges \label{alphatest}}

\begin{figure}
\includegraphics[angle=0,width=10.0cm]{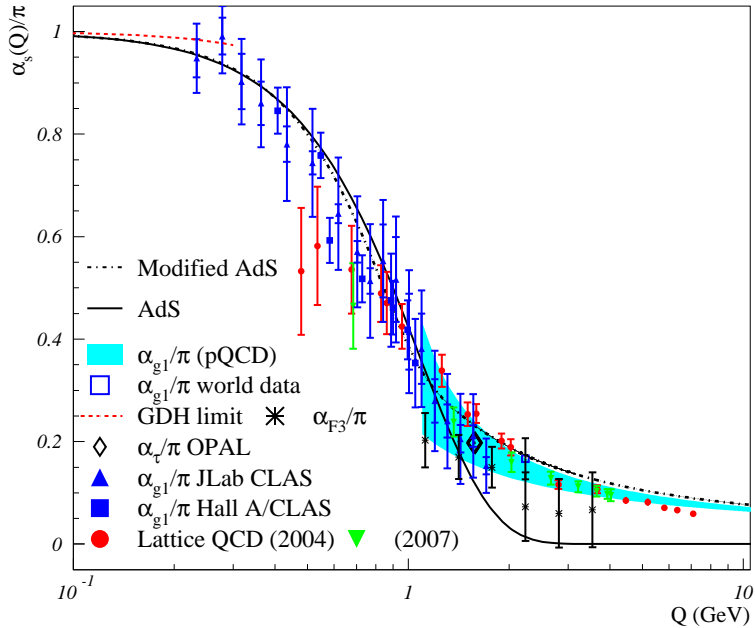}
\caption{\label{alphas}The effective coupling from LF holographic mapping  for  $\kappa = 0.54 ~ {\rm GeV}$ is compared with effective QCD couplings  extracted from
different observables and lattice results. Details on the comparison with other effective charges are given in Ref. ~\cite{Deur:2005cf}.}
\end{figure}

The effective coupling  $\alpha^{AdS}(Q^2)$ (solid line) is compared in Fig. \ref{alphas} with  experimental and lattice data. For this comparison to be meaningful, we have to impose the same normalization on the AdS coupling as the $g_1$ coupling. This defines $\alpha_s^{AdS}$ normalized to the $g_1$ scheme
\begin{equation} \label{eq:g1norm}
\alpha_{g_1}^{AdS}\left(Q^2 \! =0\right) = \pi. 
\end{equation}

A similar value for the normalization constant is derived in Ref. \cite{Erlich:2005qh} from the AdS/CFT prediction for the current-current correlator. The value of the five-dimensional coupling found in \cite{Erlich:2005qh}  for a $SU(2) $ flavor gauge theory  is $(g_5^2)_{SU(2)}  = 12 \pi^2 R /N_C$, and thus 
$(\frac{g_5^2}{4 \pi})_{SU(2)} = \pi$ for $N_C = 3$ in units $R=1$.

The couplings in Fig. \ref{alphas} agree well in the strong coupling regime  up to $Q  \! \sim \! 1$ GeV.  The value $\kappa \! = \! 0.54 ~ {\rm GeV}$ has been determined from the vector meson principal Regge trajectory.~\cite {deTeramond:2009xk}
The lattice results shown in Fig. \ref{alphas} from Ref.~\cite{Furui} have been scaled to match the perturbative UV domain. The effective charge $\alpha_{ g_1}$ has been determined in Ref~\cite{Deur:2005cf} from several experiments.  Figure \ref{alphas} also displays other couplings from different observables as well as $\alpha_{g_1}$, which is computed from the 
Bjorken sum rule~\cite{Bjorken SR}  over a large range of momentum transfer (continuous band). At $Q^2=0$ one has the constraint  on the slope of $\alpha_{g_1}$ from the Gerasimov-Drell-Hearn (GDH) sum rule~\cite{GDH}, which is also shown in the figure. 
The results show no sign of a phase transition, cusp, or other nonanalytical behavior, a fact which allows us to extend the functional dependence of the coupling to large distances. The smooth behavior of the holographic strong coupling
also allows us to extrapolate its form to the perturbative domain. This is discussed  further  in Sec. \ref{csr}.

The hadronic model obtained from the dilaton-modified AdS space provides a semiclassical first approximation to QCD.  Color confinement is introduced by the harmonic oscillator potential, but effects from gluon creation and absorption are not included in this effective theory.  The nonperturbative  confining effects vanish exponentially at large momentum transfer 
[Eq. (\ref{eq:alphaAdS})],  and thus the logarithmic falloff from pQCD quantum loops will dominate in this regime.

It is interesting to  illustrate what one expects in an augmented model which contains the standard pQCD contributions. We can use the similarity of the AdS  coupling to the effective charge  $\alpha_{g_{1}}$ at small scales as guide on how to join the perturbative and nonperturbative regimes. The fit to the data
$\alpha_{g_{1}}^{fit}$ from Ref.~\cite{Deur:2005cf}  agrees with pQCD at high momentum.   Thus, the $\alpha_{g_{1}}(Q^2)$  coupling 
provides a guide for the analytic form of the coupling over all $Q^2$. We write
\begin{equation}
\label{eq:alphafit}
\alpha_{Modified, g_1}^{AdS}(Q^2) = \alpha_{g_1}^{AdS}(Q^2) g_+(Q^2 ) + \alpha_{g_1}^{fit}(Q^2) g_-(Q^2).
\end{equation}
Here $\alpha_{g_1}^{AdS}$ is given by Eq. (\ref{eq:alphaAdS}) with the
normalization (\ref{eq:g1norm})  [continuous line in Fig.~\ref{alphas}]
and $\alpha_{g_{1}}^{fit}$ is the analytical fit to the measured coupling
$\alpha_{g_1}$.~\cite{Deur:2005cf} These couplings have the same
normalization at $Q^2=0$, given by Eq. (\ref{eq:g1norm}). We use the fit
from~\cite{Deur:2005cf} rather than using pQCD directly since
the perturbative results are meaningless near or below the transition
region and thus would not allow us to obtain a smooth transition and
analytical expression of $\alpha_{g_1}$. In order to smoothly connect the two
contributions  (dot-dashed line in Fig. \ref{alphas}), we  employ smeared step functions. For convenience we
have chosen $g_{\pm}(Q^2) = 1/(1+e^{\pm \left(Q^2 - Q^2_0\right)/\tau^2})$ with
the parameters $Q_{0}^{2}=0.8$ GeV$^{2}$ and $\tau^2=0.3$ GeV$^{2}$.

\section{Holographic $\beta$ function \label{betaAdS}}

\begin{figure}
\includegraphics[angle=0,width=10.0cm]{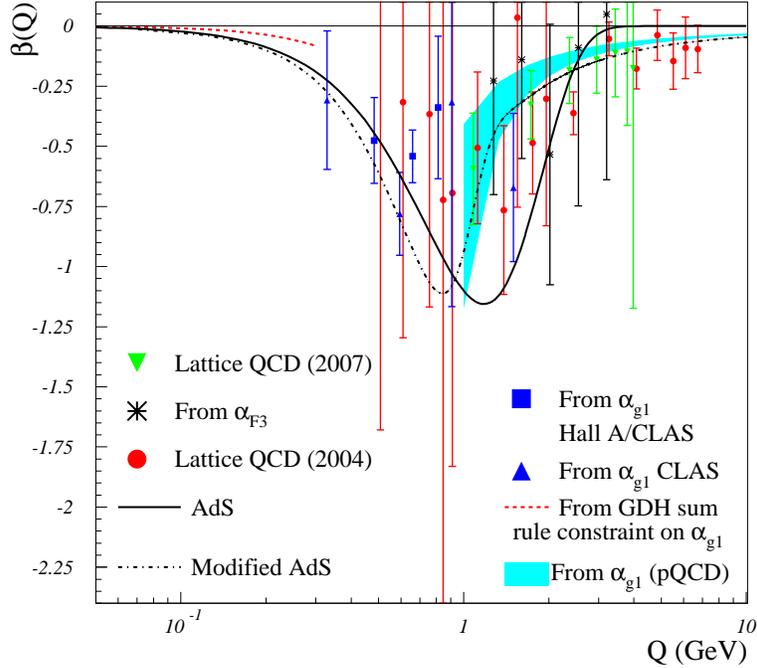}
\caption{\label{betas}Holographic model prediction for the $\beta$ function compared to  JLab
and CCFR data, lattice simulations and  results  from the Bjorken sum rule.} 
\end{figure}

The $\beta$ function for the nonperturbative  effective coupling obtained from the LF holographic mapping in a positive dilaton-modified AdS background  is
\begin{equation} \label{eq:beta}
\beta_{g_1}^{AdS}(Q^2)  = {d \over d \log{Q^2}}\alpha_{g_1}^{AdS}(Q^2) = {\pi Q^2\over 4 \kappa^2} e^{-Q^2/(4 \kappa^2)}.
\end{equation}
The solid line in Fig. \ref{betas} corresponds to the light-front holographic result Eq.  (\ref{eq:beta}).    Near $Q_0 \simeq 2 \kappa \simeq 1$ GeV, we can interpret the results as a transition from  the nonperturbative IR domain to the quark and gluon degrees of freedom in the perturbative UV  regime. The transition momentum  scale $Q_0$  is compatible with the momentum transfer for the onset of scaling behavior in exclusive reactions where quark counting rules are observed.~\cite{Brodsky:1973kr}
For example, in deuteron photo-disintegration the onset of scaling corresponds to  momentum transfer  of  1.0  GeV to the nucleon involved.~\cite{Gao:2004zh}  Dimensional counting is built into the AdS/QCD soft and hard-wall models, since the AdS amplitudes $\Phi(z)$ are governed by their twist scaling behavior $z^\tau$ at short distances, $ z \to 0$.~\cite{Polchinski:2001tt}
A similar scale for parton-hadron transition region has been observed in inclusive 
reactions.~\cite{Chen:2005tda}

Also shown on Fig. \ref{betas} are the $\beta$ functions obtained from phenomenology and lattice calculations. For clarity, we present on Fig.  \ref{betas}  only the LF holographic predictions, the lattice results from,  \cite{Furui} and the
experimental data supplemented by the relevant sum rules.  
The width of the  continuous band is computed from the uncertainty of $\alpha_{g_1}$ 
in the perturbative regime.
The dot-dashed curve corresponds to the 
extrapolated approximation given by Eq. (\ref{eq:alphafit}). Only the point-to-point uncorrelated
uncertainties  of the JLab data are used to estimate the uncertainties,  since a systematic shift cancels 
in the derivative in (\ref{eq:beta}). The data have been recombined in fewer points
to improve the statistical uncertainty;
nevertheless, the uncertainties are still large.  Upcoming JLab Hall
A and Hall B data~\cite{Upcoming JLab data} should reduce further this uncertainty.
The $\beta$ function extracted from LF holography, as well as the forms obtained from
the works of Cornwall, Bloch, Fisher {\it et al.},~\cite{S-D eq.} Burkert and Ioffe
\cite{Burkert-Ioffe} and Furui and Nakajima,~\cite{Furui}  are seen to have a similar shape
and magnitude.

Judging from these results, we infer that the   actual  $\beta$ function of QCD will interpolate between the nonperturbative results for $Q < 1$ GeV and the pQCD results 
for $Q > 1$ GeV. We also observe that the general conditions
\begin{eqnarray}
& \beta(Q \to 0) =  \beta(Q \to \infty) = 0 , \label{a} \\
&  \beta(Q)  <  0, ~ {\rm for} ~  Q > 0 , \label{b}\\
& \frac{d \beta}{d Q} \big \vert_{Q = Q_0}  = 0, \label{c} \\
& \frac{d \beta}{d Q}   < 0, ~ {\rm for} ~ Q < Q_0, ~~
 \frac{d \beta}{d Q}   > 0, ~ {\rm for} ~ Q > Q_0 \label{d} .
\end{eqnarray}
are satisfied by our model $\beta$ function obtained from LF holography.

Equation (\ref{a}) expresses the fact that  QCD approaches a conformal theory in both the far ultraviolet and deep infrared regions. In the semiclassical approximation to QCD  without particle creation or absorption,
the $\beta$ function is zero, and the approximate theory is scale  invariant
in the limit of massless quarks.~\cite{Parisi:1972zy} When quantum corrections are included,
the conformal behavior is
preserved at very large $Q$ because of asymptotic freedom and near $Q \to 0$ because the theory develops a fixed 
point.  An infrared fixed point is in fact a natural consequence of color confinement:~\cite{Cornwall:1981zr}
since the propagators of the colored fields have a maximum wavelength,  all loop
integrals in the computation of  the gluon self-energy decouple at $Q^2 \to 0$.~\cite{Brodsky:2008be} Condition (\ref{b}) for large $Q^2$, expresses the basic antiscreening behavior of QCD where the strong coupling vanishes. The $\beta$ function in QCD is essentially negative, thus the coupling increases monotonically from the UV to the IR where it reaches its maximum value:  it has a finite value for a theory with a mass gap. Equation (\ref{c}) defines the transition region at $Q_0$ where the $\beta$ function has a minimum.  Since there is only one hadronic-partonic transition, the minimum is an absolute minimum; thus the additional conditions expressed in Eq. (\ref{d}) follow immediately from 
Eqs.  (\ref{a}-\ref{c}). The conditions given by Eqs.  (\ref{a}-\ref{d}) describe the essential
behavior of the full $\beta$ function for an effective QCD coupling whose scheme/definition is similar to that of the $V$ scheme.

\section{Effective Charges and Commensurate Scale Relations \label{csr}}

As noted by Grunberg, one can use  observables such as heavy quark scattering or the Bjorken sum rule to  define effective charges $\alpha_{\cal O}(Q^2)$ each with its own physical scale.~\cite{Grunberg}   This generalizes the convention in QED where the Gell Mann-Low coupling~\cite{GellMann:1954fq} $\alpha_{QED}(Q^2)$
is defined at all scales from the scattering of infinitively heavy charged particles. Since physical  quantities are involved, the relation between effective charges cannot depend on theoretical conventions such as the
of the choice of an intermediate scheme.~\cite{Brodsky:1992pq} This is formally the transitivity
property of the renormalization group: $A$ to $B$ and $ B$ to $C$ relates $ A$
to $C$, independent of the choice of the intermediate scheme $ B$.

Although the perturbative $\beta$ function for every effective charge~\cite{Grunberg}  is universal up to two loops at high $Q^2$, each effective charge has specific characteristics, which influence its behavior at small $Q^2$.
For  example, the value and derivative of the $\alpha_{g_1}$ coupling  at $Q^2=0$ are both constrained since the Bjorken sum vanishes at $Q^2=0,$ and its derivative is given by the  GDH sum rule.~\cite{Deur:2005cf, GDH}

The relations between effective charges in pQCD are given by commensurate scale relations ~\cite{CSR}. The relative  factor between the scales of the two effective charges in the  CSR is set  to ensure that  the onset of a new quark pair in the 
$\beta$ function of the two couplings is synchronized.   This factor can be  determined by the Brodsky-Lepage-Mackenzie procedure,~\cite{Brodsky:1982gc}  where all $n_F$ and $\beta$-dependent nonconformal terms in the perturbative expansion are absorbed by the choice of the renormalization scale of the effective coupling.

This procedure also eliminates the factorial renormalon growth of perturbation theory. 
The commensurate scale relation between $\alpha_{g_1}(Q^2)$ and the Adler function effective charge $\alpha_D(Q^2)$  which is defined from $R_{e^+ e^-}$ data is now known to four loops in pQCD~\cite{Baikov:2010je}.
The relation between observables given by the CSR is independent of the choice of the intermediate renormalization scheme.
CSR are  thus precise predictions of QCD without scale or scheme ambiguity; they thus provide essential tests of the validity of QCD.

The holographic coupling  $\alpha_s^{AdS}(Q^2)$ could be seen as the nonperturbative extension of the 
$\alpha_V$ effective charge defined by Appelquist {\it et al.},~\cite{Appelquist:1977tw} and it thus can be compared to phenomenological models for the heavy quark potential such as the Cornell potential~\cite{Cornell} and lattice computations.  
Thus, an important question is how to extend the relations between observables and their effective charges to the nonperturbative domain. We can also use the CSR concept to understand the relation of $\alpha^{AdS}(Q^2)$  given by Eqs. (\ref{eq:alphaAdS}) and (\ref{eq:alphafit})  to well-measured effective charges such as the $\alpha_{g_1}$ coupling even in the nonperturbative domain.

\section{Experimental Results, Schemes  and Data Normalization \label{exp}}

The effective charges $\alpha_{g_1}$ 
and $\alpha_{F_3}$ shown in Figs. \ref{alphas} and \ref{betas} are extracted in Ref. \cite{Deur:2005cf} following the prescription of Grunberg.~\cite{Grunberg} 
Data on the spin structure function $g_{1}$, from JLab~\cite{BJ Sum data} are used to form $\alpha_{g_{1}}$.
CCFR data on the structure function $F_{3}$~\cite{CCFR}
are used to form $\alpha_{F_{3}}$, which is then related to $\alpha_{g_{1}}$
using a CSR.  
The GDH and Bjorken sum rules constrain, respectively, the small~\cite{GDH} and large~\cite{Bjorken SR} $Q^{2}$ behavior 
of the integral of  $g_{1}$ and provide a description of
$\alpha_{g_{1}}$ over a large domain. 

We note that the works
of~\cite{Grunberg} and \cite{CSR} pertain to the UV domain, whereas Ref. \cite{Deur:2005cf}
extends them to the IR region based on 
the analytical behavior of the coupling. The effective charge $\alpha_{g_1}$ is found to be approximately scale invariant in the IR domain,  in agreement with an  IR fixed point behavior.~\cite{Brodsky:2008be}
The shape of the  coupling  $\alpha_{g_1}$ agrees with other predictions of  the running coupling 
$\alpha_{s}$ at small
$Q^{2}$, including lattice QCD,~\cite{Furui} the Schwinger-Dyson
formalism,~\cite{Cornwall:1981zr, S-D eq.} and the coupling of a constituent
quark model which is consistent with hadron spectroscopy.~\cite{CQM}
We point out that the essential difference between these running couplings
is their value at $Q^2 = 0$: if normalized to the same point at
$Q^2 =0$, their $Q^2$ dependences agree within their relative uncertainties.~\cite{Deur:2005cf}

The   continuous band in Fig. 1 for $\alpha_{g_1}$ is computed with the Bjorken sum rule using the relation 
between $\alpha_{g_1}$ and $\alpha_{\overline{MS}}$.~\cite{Deur:2005cf, CSR}
The pQCD leading-twist expression of the Bjorken sum  up to third order in $\alpha_{\overline{MS}}$ is used to 
estimate the Bjorken sum. The sum rule is then used to extract $\alpha_{g_1}$ at large Q. In the pQCD 
expression of the Bjorken sum rule, 
$\alpha_{\overline{MS}}$ is retained up to second order in $\beta$ (i.e. up to $\beta_2$). The uncertainty in the  band comes from the uncertainty on $\Lambda_{\overline{MS}}=0.37^{+0.04}_{-0.07}$ and the truncation of the series.~\cite{lambda}

Although the effective coupling $\alpha_{g_1}(Q^2)$ has specific
features of deep inelastic  lepton-proton scattering, it nevertheless appears to closely mimic
the shape and magnitude of the AdS/QCD coupling near the transition region $Q \simeq Q_0$. In particular,  it
illustrates how one can have a coupling which flows analytically from
the IR strong coupling domain with an IR fixed point to the UV domain
controlled by pQCD.~\cite{CSRMS}

\begin{figure}[htbp]
\begin{center}
\includegraphics[angle=0,width=9.0cm]{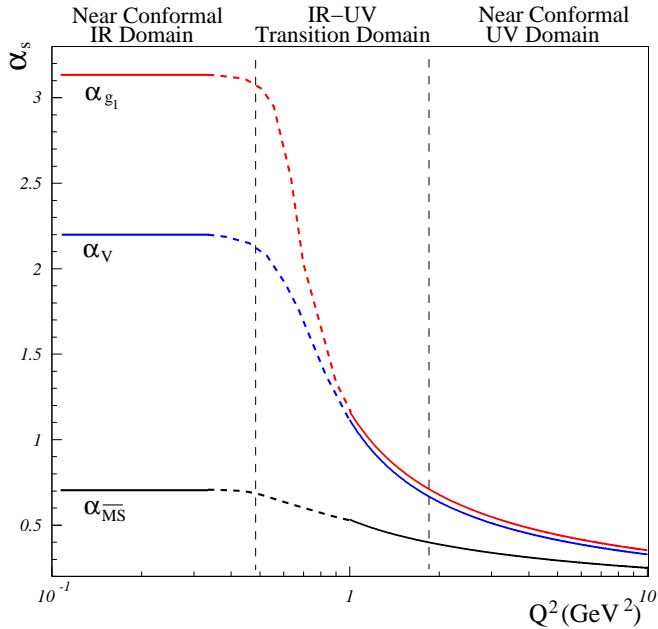}
\caption{How different schemes can lead to different values for the IR fixed point. The couplings
are computed in the UV region. They freeze in the IR region. The interpolation between UV and IR is drawn freely and is meant to be illustrative, as are the various IR fixed point values. We note that the $V$ and $g_1$ schemes are numerically close.}
\label{Fig:IR values}
\end{center}
\end{figure}

The value of $\alpha_s^{AdS}(Q)$ at $Q=0$ was not determined by our holographic approach.~\cite{size}
It is also well known that even in the pQCD domain the value of running coupling is significantly scheme dependent 
when the momentum transfer becomes small. It is  thus reasonable to assume that 
such differences propagate in the IR domain and consequently the IR value of different effective charges can differ. 
Such differences between  schemes can naturally explain the smaller IR fixed point values obtained in other computations
of the strong coupling, {\it e.g.,} in 
Ref. \cite{Cornwall:1981zr} , as qualitatively illustrated on Fig. ~\ref{Fig:IR values}.

Despite the different physics underlying the light-front holographic  coupling $\alpha_s^{AdS} (Q^2)$ and  the effective charge  $\alpha_{g_1}(Q^2)$ determined empirically from measurements of the Bjorken sum,
the shapes of the two running couplings are remarkably
close in the infrared regime.  The resemblance of $\alpha_s^{AdS} $ and  $\alpha_{g_1}$ is understandable if we recall that $\alpha_s^{AdS}$ is
a natural nonperturbative extension of $\alpha_V$.   
The scale shift in the CSR between $\alpha_V$ and $\alpha_{g_1}$  is small, making
them numerically very close.   Furthermore every effective charge 
satisfies the same pQCD $\beta$ function to two loops. Thus, the extended $\alpha_s^{AdS} $ and $\alpha_{g_1}$
are also very close at high scales.   The AdS and $g_1$ couplings
share other common features:  their $\beta$ functions have similar
structures: zero in the IR,
strongly negative in the GeV domain, and  zero  in the far UV.  We can 
exploit all of these similarities to fix the normalization $\alpha_s^{AdS}(Q=0)=\pi$ and to consistently extend the AdS
coupling to the UV domain, consistent with pQCD.

\section{Holographic Coupling in Configuration Space \label{alphar}}

In order to obtain modifications to the instantaneous  Coulomb potential 
in configuration space $V(r) = - C_F \alpha_V(r)/r$ 
from the running coupling,  one must transform the coupling defined by the static quark potential 
$V(q) = - 4 \pi C_F \alpha_V(q)/q^2$ in the nonrelativistic limit and extract the coefficient of $1/r$ to define 
the coupling $\alpha_V(r)$ in the $V$ scheme. The couplings are related by the Fourier transform~\cite{Badalian:2000hv}
\begin{equation} \label{eq:3dimFT}
\alpha_V(r) =  \frac{2}{\pi} \int_0^\infty \! dq \, \alpha_V(q) \frac{\sin(qr)}{q} .
\end{equation}
From  (\ref{eq:alphaAdS}) we find the expression 
\begin{equation}\label{eq:FTads}
\alpha_V^{AdS}(r) =  C \, {\rm{erf}}(\kappa r) = \frac{2}{ \sqrt{\pi}} \, C \int_0^{\kappa r} e^{-t^2} dt,
\end{equation}
 where $C =  \alpha_V(Q = 0) = \alpha_V(r \to \infty)$ since $\rm{erf}(x \to +\infty) = 1$.
 We have written explicitly the normalization at $Q=0$ in the $V$ scheme since it is not
expected to be equal to the normalization in the $g_1$ scheme for the reasons discussed in Sec. \ref{exp}. 

\begin{figure}[htbp]
\begin{center}
\includegraphics[angle=0,width=8.0cm]{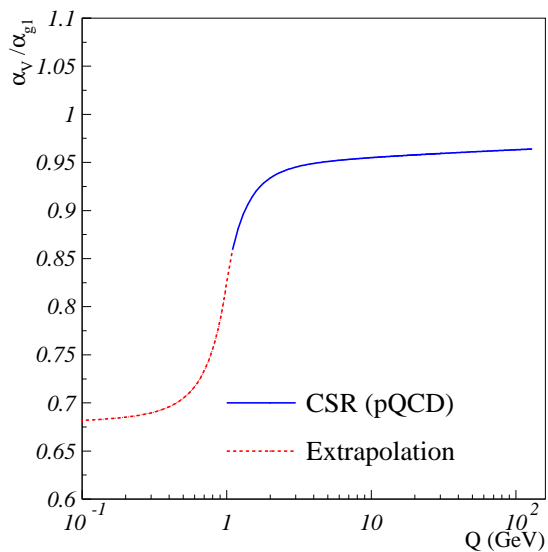}
\caption{\label{Fig:V/g} Ratio $\alpha_{V}(Q)/\alpha_{g_{1}}(Q)$. The continuous line represents the domain where the
CSR are computed at leading twist [Eq. (\ref{eq:CSR})]. The dashed line
is the extrapolation to the nonperturbative domain using the fixed point IR conformal behavior of QCD.}
\end{center}
\end{figure}

The couplings in the $V$ and $g_1$ schemes are related at leading twist by the CSR: \cite{CSR}
%\begin{multline}
\begin{equation}
\frac{\alpha_{V}(Q^{2})}{\pi}  = 
\frac{\alpha_{g_{1}}(Q^{*2})}{\pi} 
-1.09\left(\frac{\alpha_{g_{1}}(Q^{**2})}{\pi}\right)^{2}
+25.6\left(\frac{\alpha_{g_{1}}(Q^{**2})}{\pi}\right)^{3} + \cdots ,
\label{eq:CSR}
\end{equation}
%\end{multline}
with $Q^{*} = 1.18 \, Q$, $Q^{**} = 2.73 \, Q$, and we set $Q^{***}=Q^{**}$. We have verified that this 
relation numerically holds at least down to $Q^{2}=0.6$ GeV$^2$, as shown in the figure in the Appendix  (Fig.~\ref{Vg1Q}). 
In order to transform 
$\alpha_{g_{1}}(Q^2)$ into $\alpha_{V}(Q^2)$ over the full $Q^2$ range, we 
extrapolate the CSR to the nonperturbative domain. For guidance, we use the fact that QCD is near conformal
at very small $Q$; thus, the ratio $\alpha_V/\alpha_{g_{1}}$ is $Q$ independent.
A model for the ratio $\alpha_V(Q)/\alpha_{g_{1}}(Q)$  is shown in Fig. 
\ref{Fig:V/g}. We apply this ratio to $\alpha^{AdS}_{Modified,g_1}(Q)$, Eq.  (\ref{eq:alphafit}), and then Fourier 
transform the result using Eq. (\ref{eq:3dimFT}) to obtain $\alpha_{Modified,V}^{AdS}(r)$. We find $C \simeq 2.2$.

\subsection{Comparison of $V$ and $g_1$ Results}

\begin{figure}[htbp]
\begin{center}
\includegraphics[angle=0,width=12.6cm]{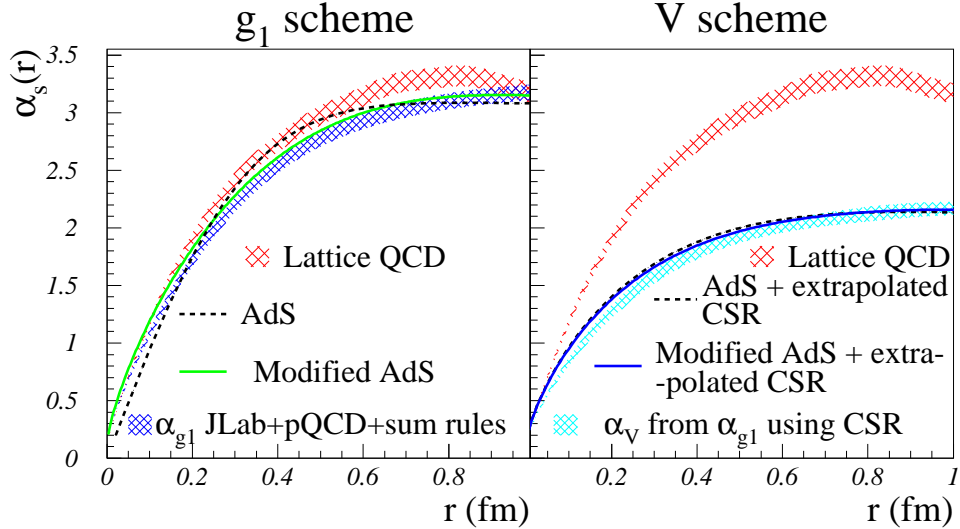}
\caption{Holographic model predictions for $\alpha_s(r)$ in configuration space in the $g_1$ scheme (left panel) and $V$ scheme (right panel).  The dashed lines are  the holographic AdS results, the continuous lines correspond to the modified
holographic results from Eq. (\ref{eq:alphafit}) normalized, respectively, to  $\alpha_{g_1}$ and $\alpha_V$ at $Q=0$.
The lower bands (cross-hatched pattern)  correspond to the JLab data and the higher (sparser pattern) to lattice results.} 
\label{D3}
\end{center}
\end{figure}

The right panel of Fig. \ref{D3} displays  $\alpha_V^{AdS}(r)$ (dashed line) and $\alpha_{V}(r)$ obtained 
with the same procedure but applied to the JLab data (lower cross-hatched band). Also shown for comparison are, on the left panel, the results 
in the $g_1$ scheme:  $\alpha_{g_1}(r)$
from JLab data (lower cross-hatched band), the light-front holographic  result from Eqs. \ref{eq:alphafit} and~\ref{eq:FTads} (continuous line) 
and lattice results from \cite{Furui} (upper cross-hatched band). The same scales are used on both panels. The fact that 
different schemes imply different values for the IR fixed point of $\alpha_s$ is exemplified in this figure in 
which $\alpha_s(r)$ in
the $V$ scheme and in the $g_1$ scheme freeze to the IR fixed point values of
$\alpha_V(Q=0)= 2.2$  and $\alpha_{g_1}(Q=0) = \pi$ respectively.

The width of the lower band on the right hand panel is the combined uncertainty on $\alpha_V$ coming from: a) the uncertainty in the value
$\Lambda_{\rm QCD}$, b) the truncation of the pQCD $\beta$-series used to 
calculate $\alpha_{\overline{MS}}$ in the perturbative region, c) the truncation of the pQCD CSR at 
$Q^{***}$ which, has been estimated by using the difference between the $Q^{**}$ and 
$Q^{***}$ orders and d) the experimental uncertainties on the 
JLab data for $\alpha_{g_{1}}$. The uncertainty coming from 
the truncation of the pQCD series for the Bjorken sum rule is negligible.

The  experimental results for $\alpha_{g_1}(r)$ follow from the integrated
JLab data according to Eq. (\ref{eq:3dimFT}). The contributions to the
integral from the unmeasured low $Q$ ($Q<0.23$ GeV) and high $Q$
($Q>1.71$ GeV) regions are computed using the sum rules \cite{GDH}
and \cite{Bjorken SR} respectively. The total experimental uncertainties, as well as the uncertainty on the large $Q$ region, are added in quadrature. This underestimates somewhat the final
uncertainty. Since $\alpha_{g_{1}}(r)$ can be computed for any
$r$, the experimental data and lattice results now appear as  bands on Fig.  \ref{D3}\,
rather than a set of data points.

\subsection{Contribution to the Instantaneous Quark-Antiquark Potential}

\begin{figure}[htbp]
\begin{center}
\includegraphics[angle=0,width=10.8cm]{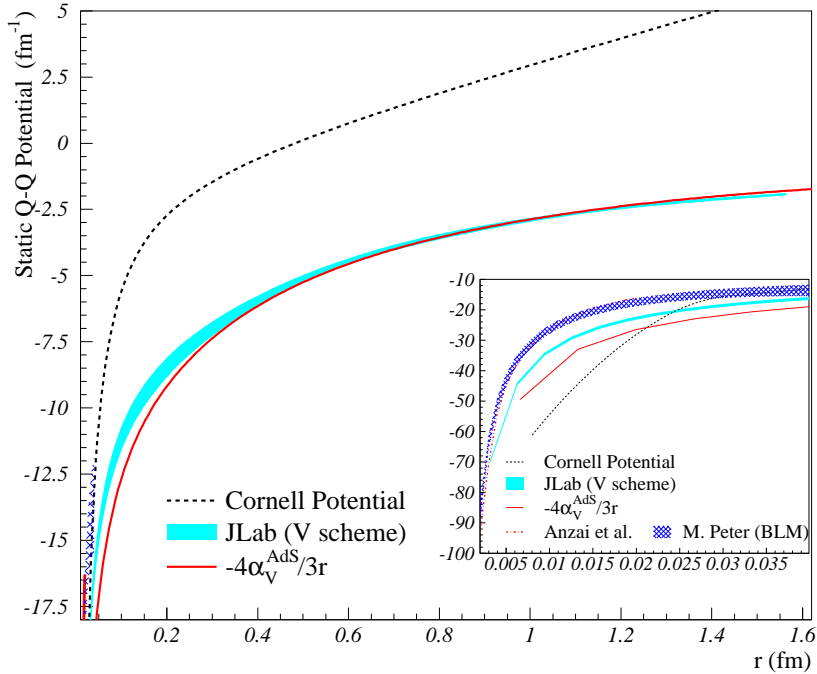}
\caption{Contribution of the running coupling to the  quark-antiquark Coulomb potential. The continuous line represents the result from light-front holography modified for pQCD effects using Eq.  (\ref{eq:alphafit}) and transformed to the $V$ scheme using the extrapolated CSR results shown in Fig. \ref{Fig:V/g}. The  continuous band is the
JLab results transformed  to the $V$ scheme using the same CSR. The dashed line is the Cornell potential.
The inserted figure zooms into the deep UV domain. PQCD results at two loops (Peter,  Ref. \cite{Peter:1997me}) and three loops (Anzai {\it et al.},  Ref. \cite{Anzai:2009tm}) are shown, respectively, by the cross-hatched band and the dot-dashed line. An arbitrary offset is applied to the Anzai {\it et al.} results.}    
\label{potential}
\end{center}
\end{figure}

The  quark-antiquark Coulomb potential 
$V(r)=-4\alpha_V(r)/3r$ is shown in Fig.~\ref{potential}  for the running coupling computed from light-front holography  and the  JLab $g_1$ measurement. The results can be compared at large distances to the phenomenological Cornell potential~\cite{Cornell} and, in the deep UV region, to the two-loop calculation of Peter~\cite{Peter:1997me} as well as with the three-loop calculation of 
Anzai {\it et al.} \cite{Anzai:2009tm}.
Other recent three-loop calculations  \cite{Smirnov:2010zc} are consistent  with the results from 
Ref. \cite{Anzai:2009tm}; the central values of the three-loop parameter $a_3$ agree within 3 \%. 
The uncertainty in Peter's result is mainly due to the uncertainty in  $\Lambda_{\overline{MS}}$, with negligible 
contributions from the truncations of the pQCD $\beta_{\overline{MS}}$ series and the CSR series. 
The truncation uncertainties are estimated as the values of the last known order of the series. All contributions 
to the uncertainty are added in quadrature.

In the case of heavy quarks  the light-front holographic equations reduce to a nonrelativistic Schr\"odinger
equation in configuration space with potential
\begin{equation}
V(r) = - \frac{4}{3} \frac{\alpha_V(r)}{r} + V_{conf}(r),
\end{equation}
where $V_{conf}$ for a soft-wall dilaton background is the potential for a three-dimensional harmonic oscillator,
$V_{conf} \simeq \frac{1}{2} m_{red} \omega^2 r^2$. Here $m_{red}$ is the reduced mass of the heavy $\overline{Q}\!-\!Q$ system, $m_{red} = m_Q m_{\overline Q}/(m_Q + m_{\overline Q})$,  and $\omega = \kappa^2/(m_Q + m_{\overline Q})$.
Remarkably, the explicit holographic confining potential $V_{conf}$, which is the dominant interaction for light quarks, vanishes
as the inverse of the quark mass for heavy quark masses.

For finite quark masses both contributions will appear. This will bring the effective  potential closer to the phenomenological
Cornell potential. Thus, the comparison of the Coulomb results  in Fig.  \ref{potential} with the Cornell potential  only holds in the limit of infinite quark masses.  A detailed discussion of the confining interaction,
its implication  for the study of the heavy meson mass spectrum, and other aspects of the instantaneous quark-antiquark potential will be discussed elsewhere.

\section{Conclusions \label{conclusions}}

We have shown that the light-front holographic mapping of  effective classical gravity in AdS space, modified by a positive-sign dilaton background $\exp{(+ \kappa^2 z^2)}$,  can be used to identify a nonperturbative effective coupling $\alpha_s^{AdS}(Q)$ and its $\beta$ function.  The same theory provides a very good description of the spectrum and form factors of light hadrons. Our analytical results for the effective holographic coupling provide new insights into the infrared
dynamics  and the form of the full $\beta$ function of QCD.

We also observe that the effective charge obtained from light-front holography is in very good agreement with the effective
coupling $\alpha_{g_1}$ extracted from  the Bjorken sum rule. Surprisingly, the Furui and Nakajima lattice results~\cite{Furui} also agree better overall with the $g_1$ scheme rather than the $V$ scheme as seen in Fig. \ref{D3}. Our analysis
indicates that light-front holography captures the essential dynamics of confinement, showing that it belongs to a universality class of models with built-in confinement. The holographic $\beta$ function shows  the transition from  nonperturbative to perturbative  regimes  at a momentum scale $Q \sim 1$ GeV and captures some of the essential characteristics of the full  $\beta$ function of QCD, thus giving further support to the application of the gauge/gravity duality to the confining dynamics of strongly coupled QCD.

We have made extensive use of commensurate scale relations, which allows us to relate observables in different 
schemes and regimes. In particular, we have extrapolated the CSR to extend the relation between observables  
to the nonperturbative domain. In the pQCD domain, we checked that the CSR are valid. This validity provides a fundamental check of QCD since the CSR are a central pQCD prediction  independent of theoretical conventions.

 The normalization of the QCD coupling $\alpha_s^{AdS}$ at
 $Q^2=0$  appears to be considerably higher than that suggested in 
 Ref. \cite{Binosi:2009qm}, a difference probably stemming from the different scheme choices.  
 However,  $\alpha_{g_1}(Q^2)$  has the advantage that it is the most precisely  
measured effective charge.  As we have noted, there is a  remarkable  
similarity of $\alpha_{g_1}(Q^2)$ to the nonperturbative strong coupling $\alpha_s^{AdS}(Q^2)$
obtained here except at large $Q^2$ where the contribution from quantum loops is dominant.  To extend its utility,
 we have provided an analytical expression encompassing the holographic result at low $Q^2$ and  pQCD contributions from gluon exchange  at large $Q^2$. 
The value of the confining scale of the model $\kappa$ is determined from the vector meson Regge trajectory, so our 
small $Q^2$-dependence prediction is parameter free.

There are many phenomenological applications where detailed knowledge of the QCD coupling and the renormalized gluon propagator at relatively soft momentum transfer are essential. 
This includes the rescattering (final-state and initial-state interactions), which
create the leading-twist Sivers single-spin correlations in 
semi-inclusive deep inelastic scattering,~\cite{Brodsky:2002cx, Collins:2002kn} the Boer-Mulders functions which lead to anomalous  $\cos 2 \phi$  contributions to the lepton pair angular distribution in the unpolarized Drell-Yan reaction,~\cite{Boer:2002ju} and the Sommerfeld-Sakharov-Schwinger correction to heavy quark production at threshold.~\cite{Brodsky:1995ds}
The confining AdS/QCD coupling from light-front holography can lead to a 
quantitative understanding of this factorization-breaking physics.~\cite{Collins:2007nk}

\begin{acknowledgments}

We thank A. Radyushkin for helpful, critical remarks. We also thank V. Burkert, J. Cornwall,  H.G. Dosch, J. Erlich,
P. H\"agler,  W. Korsch, J. K\"uhn, G. P. Lepage, T. Okui, 
and J. Papavassiliou for helpful comments. 
We thank S. Furui for sending us his recent lattice results. This research was supported by the Department of Energy under Contract 
No. DE--AC02--76SF00515. A.D.'s work is
supported by the U.S. Department of Energy (DOE). The Jefferson Science
Associates (JSA) operates the Thomas Jefferson National Accelerator
Facility for the DOE under Contract No. DE--AC05--84ER40150.
S.J.B. thanks the Hans Christian Andersen Academy and the CP$^3$-Origins Institute for their support at 
Southern Denmark University.

%,  SLAC--PUB--13840
%JLAB--PHY--10--1128

\end{acknowledgments}

\appendix*

\section{Consistency Check of Commensurate Scale Relations  \label{CSRtest}}

In this appendix we verify, within the  uncertainties discussed in the text, the validity of the CSR predictions in the pQCD domain.

\begin{figure}[htbp]
\begin{center}
\includegraphics[angle=0,width=10.0cm]{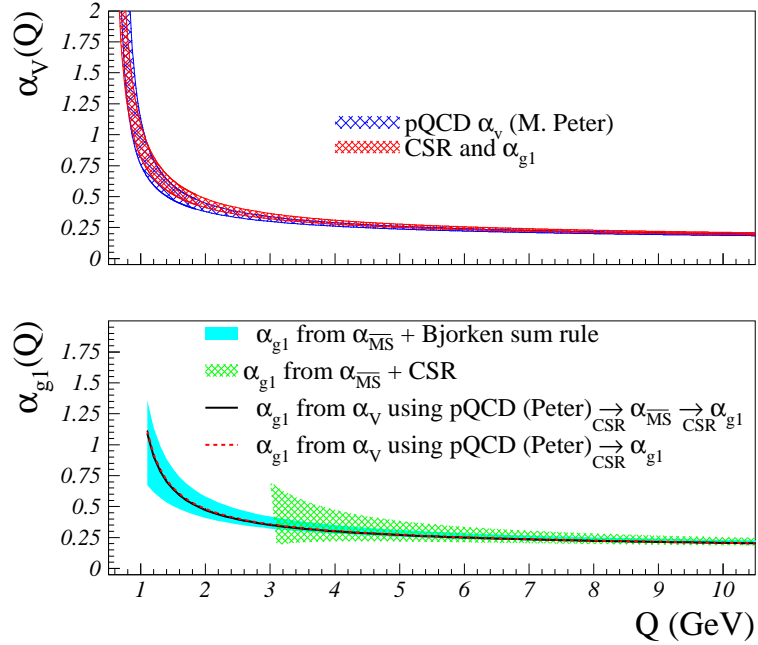}
\caption{Check of the CSR validity. Top panel: Comparison of $\alpha_V(Q^2)$ from the two-loop pQCD calculation of
Ref. \cite{Peter:1997me} and $ \alpha_V(Q^2)$ obtained using the CSR with  $\alpha_{g_1}$ as input. 
Bottom panel: Comparison of $\alpha_{g_1}(Q^2)$ computed using four different methods.
The good agreement on top and bottom panels is a fundamental check of QCD. }
\label{Vg1Q}
\end{center}
\end{figure}

The verification of CSR for different schemes is illustrated on Fig. \ref{Vg1Q}.  On the top panel of Fig. \ref{Vg1Q}  
 we compare  the full two-loop computation of $\alpha_V(Q^2)$ from Ref.~\cite{Peter:1997me} with the coupling $\alpha_V(Q^2)$ resulting from applying the 
CSR to $\alpha_{g_1}$ down to $Q^2 = 0.6 ~ {\rm GeV}^2$.  The width of the bands gives the uncertainties. For the sparse cross-hatched band (two-loop pQCD calculation), the uncertainty stems from $\Lambda_{\rm QCD}$,
 the truncation of the $\alpha_{\overline{MS}}$ series to $\beta_2$, and the truncation of the $\alpha_V$ series to
two loops ($a_2$ coefficient)  in~\cite{Peter:1997me}. All these contributions are added in quadrature. For the 
dense cross-hatched band (CSR), the uncertainties come from $\Lambda_{\rm QCD}$, the truncation of the $\alpha_{\overline{MS}}$ series to
 $\beta_2$, and the truncation of the CSR series to order $Q^{***}$.  All these contributions are again added in 
 quadrature. The various truncation uncertainties are estimated by taking the value of the last known term of the 
 series. The very good agreement of the results~\cite{uncertainties}  
allows us to check the consistency and the  applicability of CSR, even into the 
IR-UV transition region,  albeit with large uncertainties.    Throughout the paper, we limit the order of our calculation 
to $\alpha^3_s $ so that no IR terms appear.

A similar test of CSR is also shown on the bottom panel of Fig. \ref{Vg1Q}.
 It shows $\alpha_{g_1}$ computed using four different methods.
1) The continuous band corresponds to the results using the Bjorken sum rule and $\alpha_{\overline{MS}}$.
2) cross-hatched band~\cite{green band}: using the CSR to obtain $\alpha_{g_1}$ as a function of $\alpha_{\overline{MS}}$. 
3) Dashed line: using the CSR to obtain $\alpha_{g_1}$ as a function of $\alpha_V$. This latter is computed from the two-loop computation of Ref. \cite{Peter:1997me}.
4) Continuous line: using $\alpha_V$ from pQCD~\cite{Peter:1997me} as an input to
the appropriate CSR to form $\alpha_{\overline{MS}}$. This later is used as input in 
another CSR to form $\alpha_{g_1}$.
There is again excellent agreement. In addition, that the dashed and continuous lines are on top of each other verifies the transitivity property of the CSR. 
These agreements are nontrivial consistency checks of QCD  since the CSR are  central predictions of pQCD.


\begin{thebibliography}{0}

\bibitem{Binosi:2009qm} See for example:
  D.~Binosi and J.~Papavassiliou,
  ``Pinch Technique: Theory and Applications,''
  Phys.\ Rept.\  {\bf 479}, 1 (2009)
 [\href{http://arXiv.org/abs/0909.2536}{\tt arXiv:0909.2536} [hep-ph]], and references therein.
  %%CITATION = PRPLC,479,1;%%
    
    \bibitem{Maldacena:1997re}
  J.~M.~Maldacena,
 ``The large N limit of superconformal field theories and supergravity,''
  Adv.\ Theor.\ Math.\ Phys.\  {\bf 2}, 231 (1998)
  [Int.\ J.\ Theor.\ Phys.\  {\bf 38}, 1113 (1999)]
 [\href{http://arXiv.org/abs/hep-th/9711200}{\tt arXiv:hep-th/9711200}].
  %%CITATION = IJTPB,38,1113;%%
  
     \bibitem{Gubser:1998bc}
  S.~S.~Gubser, I.~R.~Klebanov and A.~M.~Polyakov,
   ``Gauge theory correlators from non-critical string theory,''
  Phys.\ Lett.\ B {\bf 428}, 105 (1998)
  [\href{http://arXiv.org/abs/hep-th/9802109}{\tt arXiv:hep-th/9802109}];
  %%CITATION = HEP-TH 9802109;%%
%\bibitem{Witten:1998qj}
  E.~Witten,
  ``Anti-de Sitter space and holography,''
  Adv.\ Theor.\ Math.\ Phys.\  {\bf 2}, 253 (1998)
  [\href{http://arXiv.org/abs/hep-th/9802150}{\tt arXiv:hep-th/9802150}].
  %%CITATION = HEP-TH 9802150;%%
  
   \bibitem{deTeramond:2008ht}
  G.~F.~de Teramond and S.~J.~Brodsky,
  ``Light-Front Holography: A First Approximation to QCD,''
  Phys.\ Rev.\ Lett.\  {\bf 102}, 081601 (2009)
 [\href{http://arXiv.org/abs/0809.4899}{\tt arXiv:0809.4899} [hep-ph]].
  %%CITATION = ARXIV:0809.4899;%%
  
  \bibitem{Brodsky:2006uqa}
  S.~J.~Brodsky and G.~F.~de Teramond,
  ``Hadronic spectra and light-front wavefunctions in holographic QCD,''
  Phys.\ Rev.\ Lett.\  {\bf 96}, 201601 (2006)
  [\href{http://arXiv.org/abs/hep-ph/0602252}{\tt arXiv:hep-ph/0602252}];
  %%CITATION = PRLTA,96,201601;%%
  % \bibitem{Brodsky:2007hb}
%  S.~J.~Brodsky and G.~F.~de Teramond,
  ``Light-Front Dynamics and AdS/QCD Correspondence: The Pion Form Factor in the Space- and timelike Regions,''
  Phys.\ Rev.\  D {\bf 77}, 056007 (2008)
 [\href{http://arXiv.org/abs/0707.3859}{\tt arXiv:0707.3859} [hep-ph]].
  %%CITATION = PHRVA,D77,056007;%%
  
    \bibitem{Brodsky:2008pf}
 S.~J.~Brodsky and G.~F.~de Teramond,
  ``Light-Front Dynamics and AdS/QCD Correspondence: Gravitational Form Factors of Composite Hadrons,''
  Phys.\ Rev.\  D {\bf 78}, 025032 (2008)
   [\href{http://arXiv.org/abs/0804.0452}{\tt arXiv:0804.0452} [hep-ph]].
  %%CITATION = PHRVA,D78,025032;%%
  
   \bibitem{deTeramond:2009xk}
  G.~F.~de Teramond and S.~J.~Brodsky,
  ``Light-Front Holography and Gauge/Gravity Duality: The Light Meson and Baryon Spectra,''
  Nucl.\ Phys.\ B, Proc. Suppl. {\bf 199}, 89 (2010)
  [\href{http://arXiv.org/abs/0909.3900}{\tt arXiv:0909.3900} [hep-ph]].
  %%CITATION = ARXIV:0909.3900;%%
 
 \bibitem{deTeramond:2010we}
  G.~F.~de Teramond and S.~J.~Brodsky,
  ``Light-Front Quantization Approach to the Gauge-Gravity Correspondence and Hadron Spectroscopy,''
  \href{http://arXiv.org/abs/1001.5193}{\tt arXiv:1001.5193} [hep-ph].
  %%CITATION = ARXIV:1001.5193;%%
  
   \bibitem{Grunberg} 
  %\bibitem{Grunberg:1980ja}
  G.~Grunberg,
  ``Renormalization Group Improved Perturbative QCD,''
  Phys.\ Lett.\  B {\bf 95}, 70 (1980);
 % [Erratum-ibid.\  B {\bf 110}, 501 (1982)];
  %%CITATION = PHLTA,B95,70;%%
   %\bibitem{Grunberg:1982fw}
 %G.~Grunberg,
 ``Renormalization Scheme Independent QCD and QED: The Method of Effective Charges,''
 Phys.\ Rev.\  D {\bf 29}, 2315 (1984);
 %%CITATION = PHRVA,D29,2315;%%
% \bibitem{Grunberg:1989xf}
% G.~Grunberg,
``On Some Ambiguities in the Method of Effective Charges",
  Phys.\ Rev.\  D {\bf 40}, 680 (1989).
  %%CITATION = PHRVA,D40,680;%%
  
   \bibitem{CSR}
%\bibitem{Brodsky:1994eh}
  S.~J.~Brodsky and H.~J.~Lu,
  ``Commensurate scale relations in quantum chromodynamics,''
  Phys.\ Rev.\  D {\bf 51}, 3652 (1995)
  [\href{http://arXiv.org/abs/hep-ph/9405218}{\tt arXiv:hep-ph/9405218}];
  %%CITATION = PHRVA,D51,3652;%%
%\bibitem{Brodsky:1995tb}
  S.~J.~Brodsky, G.~T.~Gabadadze, A.~L.~Kataev and H.~J.~Lu,
  ``The generalized Crewther relation in QCD and its experimental consequences,''
  Phys.\ Lett.\  B {\bf 372}, 133 (1996)
  [\href{http://arXiv.org/abs/hep-ph/9512367}{\tt arXiv:hep-ph/9512367}].
  %%CITATION = PHLTA,B372,133;%%
  
  \bibitem{Appelquist:1977tw}
  T.~Appelquist, M.~Dine and I.~J.~Muzinich,
 ``The Static Potential in Quantum Chromodynamics,''
  Phys.\ Lett.\  B {\bf 69}, 231 (1977);
  %%CITATION = PHLTA,B69,231;%%
%\bibitem{Appelquist:1977es}
 % T.~Appelquist, M.~Dine and I.~J.~Muzinich,
  ``The Static Limit of Quantum Chromodynamics,''
  Phys.\ Rev.\  D {\bf 17}, 2074 (1978).
  %%CITATION = PHRVA,D17,2074;%%
  
   \bibitem{Bjorken SR}
%\bibitem{Bjorken:1966jh}
  J.~D.~Bjorken,
  ``Applications of the Chiral $U(6) \otimes U(6)$ Algebra of Current Densities,''
  Phys.\ Rev.\  {\bf 148}, 1467 (1966).
  %%CITATION = PHRVA,148,1467;%%

\bibitem{Deur:2009hu}
  A.~Deur,
  ``Study of spin sum rules and the strong coupling constant at large distances,''
  \href{http://arXiv.org/abs/0907.3385}{\tt arXiv:0907.3385} [nucl-ex].
  %%CITATION = ARXIV:0907.3385;%%
  
  \bibitem{FFexample}
  One can question the relevance of fundamental quark constituents and couplings in the domain below the transition
regime since the effective degrees of freedom are hadronic. To answer this, consider  the EM form factor of a pion $F(Q^2)$ defined by the transition matrix element of the EM current between hadronic  states 
$\langle P' \vert J^\mu \vert P \rangle = (P + P') F(Q^2)$, an expression valid for any value of the momentum transfer $Q = P' - P$. Thus even if $Q$ is near zero, the EM couplings are dictated by the 
quark current $J^\mu = e_q \bar q \gamma^\mu q$.  The gluon couples in a similar way to the fundamental constituents. Thus hadronic  interactions at any scale are governed by the underlying quark and gluon dynamics,  even if an effective low energy theory that inherits the fundamental QCD symmetries also exists.  

  \bibitem{Polchinski:2001tt}
  J.~Polchinski and M.~J.~Strassler,
 ``Hard scattering and gauge/string duality,''
  Phys.\ Rev.\ Lett.\  {\bf 88}, 031601 (2002)
  [\href{http://arXiv.org/abs/hep-th/0109174}{\tt arXiv:hep-th/0109174}].
  %%CITATION = HEP-TH 0109174;%%
  
  \bibitem{Brodsky:1973kr}
  S.~J.~Brodsky and G.~R.~Farrar,
 ``Scaling Laws at Large Transverse Momentum,''
  Phys.\ Rev.\ Lett.\  {\bf 31}, 1153 (1973);
  %%CITATION = PRLTA,31,1153;%%
 %\bibitem{Matveev:ra}
  V.~A.~Matveev, R.~M.~Muradian and A.~N.~Tavkhelidze,
 ``Automodellism in the Large-Angle Elastic Scattering and Structure of Hadrons,''
  Lett.\ Nuovo Cim.\  {\bf 7}, 719 (1973).
  %%CITATION = NCLTA,7,719;%%
  
    \bibitem{Karch:2006pv}
  A.~Karch, E.~Katz, D.~T.~Son and M.~A.~Stephanov,
  ``Linear Confinement and AdS/QCD,''
  Phys.\ Rev.\  D {\bf 74}, 015005 (2006)
  [\href{http://arXiv.org/abs/hep-ph/0602229}{\tt arXiv:hep-ph/0602229}].
  %%CITATION = PHRVA,D74,015005;%%
  
  \bibitem{Andreev:2006ct}
  O.~Andreev and V.~I.~Zakharov,
  ``Heavy-quark potentials and AdS/QCD,''
  Phys.\ Rev.\  D {\bf 74}, 025023 (2006)
  [\href{http://arXiv.org/abs/hep-ph/0604204}{\tt arXiv:hep-ph/0604204}].
  %%CITATION = PHRVA,D74,025023;%%
  
   \bibitem{Glazek:1987ic}
  S.~D.~Glazek and M.~Schaden,
  ``Gluon Condensate Induces Confinement in Mesons and Baryons,"
  Phys.\ Lett.\  B {\bf 198}, 42 (1987).
  %%CITATION = PHLTA,B198,42;%%  
  
  \bibitem{Gherghetta:2009ac}
  T.~Gherghetta, J.~I.~Kapusta and T.~M.~Kelley,
  ``Chiral symmetry breaking in the soft-wall AdS/QCD model,''
  Phys.\ Rev.\  D {\bf 79} (2009) 076003
  [\href{http://arXiv.org/abs/0902.1998}{\tt arXiv:0902.1998} [hep-ph]].
  %%CITATION = PHRVA,D79,076003;%%
  
  \bibitem{Zuo:2009dz}
  F.~Zuo,
  ``Improved Soft-Wall model with a negative dilaton,''
 \href{http://arXiv.org/abs/0909.4240}{\tt arXiv:0909.4240} [hep-ph].
  %%CITATION = ARXIV:0909.4240;%%
 See also
% \bibitem{Afonin:2010fr}
  S.~S.~Afonin,
  ``Holographic models for planar QCD without AdS/CFT correspondence,''
\href{http://arXiv.org/abs/1001.3105}{\tt arXiv:1001.3105 [hep-ph]}.
  %%CITATION = ARXIV:1001.3105;%%

  \bibitem{Erlich:2005qh}
  J.~Erlich, E.~Katz, D.~T.~Son and M.~A.~Stephanov,
  ``QCD and a holographic model of hadrons,''
  Phys.\ Rev.\ Lett.\  {\bf 95}, 261602 (2005)
  [\href{http://arXiv.org/abs/hep-ph/0501128}{\tt arXiv:hep-ph/0501128}].
  %%CITATION = PRLTA,95,261602;%%
  
  \bibitem{DaRold:2005zs}
  L.~Da Rold and A.~Pomarol,
  ``Chiral symmetry breaking from five dimensional spaces,''
  Nucl.\ Phys.\ B {\bf 721}, 79 (2005)
  [\href{http://arXiv.org/abs/hep-ph/0501218}{\tt arXiv:hep-ph/0501218}];
  %%CITATION = HEP-PH 0501218;%%
%  \bibitem{DaRold:2005vr}
 % L.~Da Rold and A.~Pomarol,
 ``The scalar and pseudoscalar sector in a five-dimensional approach to chiral symmetry breaking,''
  JHEP {\bf 0601}, 157 (2006)
  [\href{http://arXiv.org/abs/hep-ph/0510268}{\tt arXiv:hep-ph/0510268}].
  %%CITATION = JHEPA,0601,157;%%
  
   \bibitem{Brodsky:2003px}
  S.~J.~Brodsky and G.~F.~de Teramond,
  ``Light-front hadron dynamics and AdS/CFT correspondence,''
  Phys.\ Lett.\  B {\bf 582}, 211 (2004)
  [\href{http://arXiv.org/abs/hep-th/0310227}{\tt arXiv:hep-th/0310227}].
  %%CITATION = PHLTA,B582,211;%%
    
   \bibitem{Dirac:1949cp}
  P.~A.~M.~Dirac,
  ``Forms of Relativistic Dynamics,''
  Rev.\ Mod.\ Phys.\  {\bf 21}, 392 (1949).
  %%CITATION = RMPHA,21,392;%%
    
\bibitem{Brodsky:1997de}
 S.~J.~Brodsky, H.~C.~Pauli and S.~S.~Pinsky,
 ``Quantum Chromodynamics and Other Field Theories on the Light Cone,''
 Phys.\ Rept.\  {\bf 301}, 299 (1998)
 [\href{http://arXiv.org/abs/hep-ph/9705477}{\tt arXiv:hep-ph/9705477}].
 %%CITATION = PRPLC,301,299;%%
 
  \bibitem{Soper:1976jc}
  D.~E.~Soper,
  ``The Parton Model and the Bethe-Salpeter Wave Function,''
  Phys.\ Rev.\ D {\bf 15}, 1141 (1977).
  %%CITATION = PHRVA,D15,1141;%%
  
   \bibitem{Cornwall:1981zr}
  J.~M.~Cornwall,
  ``Dynamical Mass Generation in Continuum QCD,''
  Phys.\ Rev.\  D {\bf 26}, 1453 (1982).
  %%CITATION = PHRVA,D26,1453;%%

\bibitem{Brodsky:2008be}
 S.~J.~Brodsky and R.~Shrock,
``Maximum Wavelength of Confined Quarks and Gluons and Properties of Quantum Chromodynamics,''
 Phys.\ Lett.\  B {\bf 666}, 95 (2008)
[\href{http://arXiv.org/abs/0806.1535}{\tt arXiv:0806.1535} [hep-th]].
 %%CITATION = PHLTA,B666,95;%%
   
 
 \bibitem{Csaki:2006ji}
  C.~Csaki and M.~Reece,
 ``Toward a systematic holographic QCD: A braneless approach,''
  JHEP {\bf 0705}, 062 (2007)
  [\href{http://arXiv.org/abs/hep-ph/0608266}{\tt arXiv:hep-ph/0608266}].
  %%CITATION = JHEPA,0705,062;%%
  
   \bibitem{Hirn:2005vk}
  J.~Hirn, N.~Rius and V.~Sanz,
  ``Geometric approach to condensates in holographic QCD,''
  Phys.\ Rev.\  D {\bf 73}, 085005 (2006)
  [\href{http://arXiv.org/abs/hep-ph/0512240}{\tt arXiv:hep-ph/0512240}].
  %%CITATION = PHRVA,D73,085005;%%
    
  \bibitem{Gursoy:2007cb}
  U.~Gursoy and E.~Kiritsis,
  ``Exploring improved holographic theories for QCD: Part I,''
  JHEP {\bf 0802}, 032 (2008)
  [\href{http://arXiv.org/abs/0707.1324}{\tt arXiv:0707.1324} [hep-th]];
  %%CITATION = JHEPA,0802,032;%%
 % \bibitem{Gursoy:2007er}
  U.~Gursoy, E.~Kiritsis and F.~Nitti,
  ``Exploring improved holographic theories for QCD: Part II,''
  JHEP {\bf 0802}, 019 (2008)
  [\href{http://arXiv.org/abs/0707.1349}{\tt arXiv:0707.1349} [hep-th]].
  %%CITATION = JHEPA,0802,019;%%
  
 
\bibitem{Zeng:2008sx}
  D.~F.~Zeng,
  ``Heavy quark potentials in some renormalization group revised AdS/QCD models,''
  Phys.\ Rev.\  D {\bf 78}, 126006 (2008)
  [\href{http://arXiv.org/abs/0805.2733}{\tt arXiv:0805.2733} [hep-th]].
  %%CITATION = PHRVA,D78,126006;%%
    
   \bibitem{Galow:2009kw}
  B.~Galow, E.~Megias, J.~Nian and H.~J.~Pirner,
 ``Phenomenology of AdS/QCD and its Gravity Dual,''
 Nucl.\ Phys.\  B {\bf 834}, 330 (2010)
  [\href{http://arXiv.org/abs/0911.0627}{\tt arXiv:0911.0627} [hep-ph]].
  %%CITATION = NUPHA,B834,330;%%
   
  \bibitem{Alanen:2009xs}
  J.~Alanen, K.~Kajantie and V.~Suur-Uski,
  ``A gauge/gravity duality model for gauge theory thermodynamics,''
  Phys.\ Rev.\  D {\bf 80}, 126008 (2009)
  [\href{http://arXiv.org/abs/0911.2114}{\tt arXiv:0911.2114} [hep-ph]].
  %%CITATION = PHRVA,D80,126008;%%
  
  \bibitem{Jarvinen:2009fe}
  M.~Jarvinen and F.~Sannino,
  ``Holographic Conformal Window - A Bottom Up Approach,''
  JHEP {\bf 1005}, 041 (2010)
 [\href{http://arXiv.org/abs/0911.2462}{\tt arXiv:0911.2462} [hep-ph]].
  %%CITATION = JHEPA,1005,041;%%

  
 \bibitem{Peet:1998wn}
 A.~W.~Peet and J.~Polchinski,
 ``UV/IR relations in AdS dynamics,''
 Phys.\ Rev.\  D {\bf 59}, 065011 (1999)
 [\href{http://arXiv.org/abs/hep-th/9809022}{\tt arXiv:hep-th/9809022}].
 %%CITATION = PHRVA,D59,065011;%%
  
  \bibitem{Furui}
 %\bibitem{Furui:2004cx}
  S.~Furui and H.~Nakajima,
  ``What the Gribov copy tells about confinement and the theory of dynamical chiral symmetry breaking,''
  Phys.\ Rev.\  D {\bf 70}, 094504 (2004);
  %%CITATION = PHRVA,D70,094504;%%
  %\bibitem{Furui:2009nj}
  S.~Furui,
 ``Self-dual gauge fields, domain wall fermion zero modes and the Kugo-Ojima confinement criterion,''
  \href{http://arXiv.org/abs/0908.2768}{\tt arXiv:0908.2768} [hep-lat].
  %%CITATION = ARXIV:0908.2768;%%
  
   \bibitem{Deur:2005cf}
  A.~Deur, V.~Burkert, J.~P.~Chen and W.~Korsch,
 ``Experimental determination of the effective strong coupling constant,''
  Phys.\ Lett.\  B {\bf 650}, 244 (2007)
  [\href{http://arXiv.org/abs/hep-ph/0509113}{\tt arXiv:hep-ph/0509113}];
  %%CITATION = PHLTA,B650,244;%%
  %\bibitem{Deur:2008rf}
  %A.~Deur, V.~Burkert, J.~P.~Chen and W.~Korsch,
  ``Determination of the effective strong coupling constant $\alpha_{s,g_1}(Q^2)$ from CLAS spin structure function data,''
  Phys.\ Lett.\  B {\bf 665}, 349 (2008)
  [\href{http://arXiv.org/abs/0803.4119}{\tt arXiv:0803.4119} [hep-ph]].
  %%CITATION = PHLTA,B665,349;%%
  
    \bibitem{GDH}
%\bibitem{Drell:1966jv}
  S.~D.~Drell and A.~C.~Hearn,
  ``Exact Sum Rule for Nucleon Magnetic Moments,''
  Phys.\ Rev.\ Lett.\  {\bf 16}, 908 (1966);
  %%CITATION = PRLTA,16,908;%%
  %\bibitem{Gerasimov:1965et}
  S.~B.~Gerasimov,
  ``A sum rule for magnetic moments and the damping of the nucleon magnetic moment in nuclei,''
  Sov.\ J.\ Nucl.\ Phys.\  {\bf 2} (1966) 430
  [Yad.\ Fiz.\  {\bf 2} (1966) 598].
  %%CITATION = YAFIA,2,598;%%

  \bibitem{Gao:2004zh}
 See, for example,
  H.~Gao and L.~Zhu,
 ``Pion photoproduction on the nucleon,''
  AIP Conf.\ Proc.\  {\bf 747}, 179 (2005)
  [\href{http://arXiv.org/abs/nucl-ex/0411014}{\tt arXiv:nucl-ex/0411014}], 
  and references therein.
  %%CITATION = APCPC,747,179;%%
  
  \bibitem{Chen:2005tda}
  See, for example,
  J.~P.~Chen, A.~Deur and Z.~E.~Meziani,
  ``Sum rules and moments of the nucleon spin structure functions,''
  Mod.\ Phys.\ Lett.\  A {\bf 20} (2005) 2745
  [\href{http://arXiv.org/abs/nucl-ex/0509007}{\tt arXiv:nucl-ex/0509007}],
  and references therein.
  %%CITATION = MPLAE,A20,2745;%%
        
 \bibitem{Upcoming JLab data}
Upcoming JLab data: JLab experiments E97-110,
J-P Chen, A. Deur, F. Garibaldi spokespersons; EG1b, V. Burkert, D.
Crabb, M. Tauiti, G. Dodge and S. Khun spokespersons; EG4,  M. Battaglieri,
A. Deur, R. DeVita, G. Dodge, M. Ripani and K. Slifer spokespersons.

 \bibitem{S-D eq.}
 %\bibitem{Bloch:2002eq}
  J.~C.~R.~Bloch,
  ``Multiplicative renormalizability and quark propagator,''
  Phys.\ Rev.\  D {\bf 66}, 034032 (2002)
  [\href{http://arXiv.org/abs/hep-ph/0202073}{\tt arXiv:hep-ph/0202073}];
  %%CITATION = PHRVA,D66,034032;%%
  %\bibitem{Maris:1999nt}
  P.~Maris and P.~C.~Tandy,
  ``Bethe-Salpeter study of vector meson masses and decay constants,''
  Phys.\ Rev.\  C {\bf 60}, 055214 (1999)
  [\href{http://arXiv.org/abs/nucl-th/9905056}{\tt arXiv:nucl-th/9905056}];
  %%CITATION = PHRVA,C60,055214;%%
 %\ bibitem{Fischer:2002hna}
  C.~S.~Fischer and R.~Alkofer,
  ``Infrared exponents and running coupling of SU(N) Yang-Mills theories,''
  Phys.\ Lett.\  B {\bf 536}, 177 (2002)
  [\href{http://arXiv.org/abs/hep-ph/0202202}{\tt arXiv:hep-ph/0202202}];
  %%CITATION = PHLTA,B536,177;%%
 %\bibitem{Fischer:2002eq}
  C.~S.~Fischer, R.~Alkofer and H.~Reinhardt,
  ``The elusiveness of infrared critical exponents in Landau gauge  Yang-Mills theories,''
  Phys.\ Rev.\  D {\bf 65}, 094008 (2002)
  [\href{http://arXiv.org/abs/hep-ph/0202195}{\tt rXiv:hep-ph/0202195}];
  %%CITATION = PHRVA,D65,094008;%%
  %\bibitem{Alkofer:2002ne}
  R.~Alkofer, C.~S.~Fischer and L.~von Smekal,
  ``The infrared behaviour of the running coupling in Landau gauge QCD,''
  Acta Phys.\ Slov.\  {\bf 52}, 191 (2002)
  [\href{http://arXiv.org/abs/hep-ph/0205125}{\tt arXiv:hep-ph/0205125}];
  %%CITATION = APSVC,52,191;%%
 %\ bibitem{Bhagwat:2003vw}
  M.~S.~Bhagwat, M.~A.~Pichowsky, C.~D.~Roberts and P.~C.~Tandy,
 ``Analysis of a quenched lattice-QCD dressed-quark propagator,''
  Phys.\ Rev.\  C {\bf 68}, 015203 (2003)
  [\href{http://arXiv.org/abs/nucl-th/0304003}{\tt arXiv:nucl-th/0304003}].
  %%CITATION = PHRVA,C68,015203;%%

\bibitem{Burkert-Ioffe}
%\bibitem{Burkert:1992tg}
  V.~D.~Burkert and B.~L.~Ioffe,
 ``On the $Q^2$ variation of spin dependent deep inelastic electron - proton scattering,''
  Phys.\ Lett.\  B {\bf 296}, 223 (1992);
  %%CITATION = PHLTA,B296,223;%%
%\bibitem{Burkert:1993ya}
%  V.~D.~Burkert and B.~L.~Ioffe,
 ``Polarized Structure Functions of Proton and Neutron and The Gerasimov-Drell-Hearn and Bjorken Sum Rules,''
  J.\ Exp.\ Theor.\ Phys.\  {\bf 78}, 619 (1994)
  [Zh.\ Eksp.\ Teor.\ Fiz.\  {\bf 105}, 1153 (1994)].
  %%CITATION = ZETFA,105,1153;%%
  
  \bibitem{Parisi:1972zy}
  G.~Parisi,
  ``Conformal invariance in perturbation theory,''
  Phys.\ Lett.\  B {\bf 39}, 643 (1972).
  %%CITATION = PHLTA,B39,643;%%'
  
\bibitem{GellMann:1954fq}
  M.~Gell-Mann and F.~E.~Low,
 ``Quantum electrodynamics at small distances,''
  Phys.\ Rev.\  {\bf 95}, 1300 (1954).
  %%CITATION = PHRVA,95,1300;%%

\bibitem{Brodsky:1992pq}
  S.~J.~Brodsky and H.~J.~Lu,
  ``On the selfconsistency of scale setting methods,''
  \href{http://arXiv.org/abs/hep-ph/9211308}{\tt arXiv:hep-ph/9211308}.
  %%CITATION = HEP-PH/9211308;%%
  
   \bibitem{Brodsky:1982gc}
  S.~J.~Brodsky, G.~P.~Lepage and P.~B.~Mackenzie,
  ``On the Elimination of Scale Ambiguities in Perturbative Quantum Chromodynamics,''
  Phys.\ Rev.\  D {\bf 28}, 228 (1983).
  %%CITATION = PHRVA,D28,228;%%
  
  \bibitem{Baikov:2010je}
 P.~A.~Baikov, K.~G.~Chetyrkin and J.~H.~Kuhn,
 ``Adler Function, Bjorken Sum Rule, and the Crewther Relation to Order $\alpha_s^4$ in a General Gauge Theory,''
  Phys.\ Rev.\ Lett.\  {\bf 104}, 132004 (2010)
  [\href{http://arXiv.org/abs/1001.3606}{\tt arXiv:1001.3606} [hep-ph]].
  %%CITATION = PRLTA,104,132004;%%
   
  \bibitem{Cornell}
%  \bibitem{Eichten:1978tg}
  E.~Eichten, K.~Gottfried, T.~Kinoshita, K.~D.~Lane and T.~M.~Yan,
  ``Charmonium: The Model,''
  Phys.\ Rev.\  D {\bf 17}, 3090 (1978);
 % [Erratum-ibid.\  D {\bf 21}, 313 (1980)].
  %%CITATION = PHRVA,D17,3090;%%
%\bibitem{Eichten:1979ms}
%  E.~Eichten, K.~Gottfried, T.~Kinoshita, K.~D.~Lane and T.~M.~Yan,
  ``Charmonium: Comparison With Experiment,''
  Phys.\ Rev.\  D {\bf 21}, 203 (1980).
  %%CITATION = PHRVA,D21,203;%%
 
\bibitem{BJ Sum data} 
%\bibitem{Deur:2004ti}
  A.~Deur {\it et al.},
  ``Experimental determination of the evolution of the Bjorken integral at  low $Q^2$,''
  Phys.\ Rev.\ Lett.\  {\bf 93}, 212001 (2004)
 [\href{http://arXiv.org/abs/hep-ex/0407007}{\tt arXiv:hep-ex/0407007}];
  %%CITATION = PRLTA,93,212001;%%
  %\bibitem{Deur:2008ej}
  A.~Deur {\it et al.},
  ``Experimental study of isovector spin sum rules,''
  Phys.\ Rev.\  D {\bf 78}, 032001 (2008)
  [\href{http://arXiv.org/abs/nucl-ex/0802.3198}{\tt arXiv:0802.3198} [nucl-ex]].
  %%CITATION = PHRVA,D78,032001;%%
  
  \bibitem{CCFR}
%\bibitem{Kim:1998kia}
  J.~H.~Kim {\it et al.},
  ``A measurement of $\alpha_s(Q^2)$ from the Gross-Llewellyn Smith sum  rule,''
  Phys.\ Rev.\ Lett.\  {\bf 81}, 3595 (1998)
  [\href{http://arXiv.org/abs/hep-ex/9808015}{\tt arXiv:hep-ex/9808015}].
  %%CITATION = PRLTA,81,3595;%%

  \bibitem{CQM}
%\bibitem{Godfrey:1985xj}
  S.~Godfrey and N.~Isgur,
  ``Mesons in a Relativized Quark Model With Chromodynamics,''
  Phys.\ Rev.\  D {\bf 32}, 189 (1985).
  %%CITATION = PHRVA,D32,189;%%
  
  \bibitem{lambda}
  The value of $\Lambda_{\overline{MS}}$ is chosen so that
$\alpha_{\overline{MS}}(Q = $1 GeV$)=0.45 \pm 0.05$,  consistent with the
 value extracted from PDG.~\cite{PDG} The propagated $\Lambda_{\overline{MS}}$
uncertainty is added in quadrature with the uncertainties due to the
truncations of the $\beta$ function and of the pQCD Bjorken sum rule
series. The uncertainty due to the truncation of the 
$\beta$ function is taken as the difference in $\alpha_{\overline{MS}}$ computed to second and first order in $\beta$. Similarly, the uncertainty due to the truncation of the pQCD expression of the Bjorken sum rule is taken as the difference between its estimates up to $\alpha_{\overline{MS}}^2$ and $\alpha_{\overline{MS}}^3$. 

  \bibitem{PDG}
C.~Amsler {\it et al.}  [Particle Data Group],
  ``Review of Particle Physics,''
  Phys.\ Lett.\  B {\bf 667}, 1 (2008).

 
 \bibitem{CSRMS}
The CSR concept shows  that perturbation theory in the $\overline{MS}$ 
scheme is poorly convergent at small values of $Q^2$. For example, the first-order  CSR
scale shift from the $\overline{MS}$ scheme to 
the $V$ scheme is ${Q^*}^2 = 0.189 \, Q^2$,~\cite{CSR} thus reducing 
significantly the range of applicability  of pQCD. 
In contrast, the first-order scale shift 
from the $g_1$ scheme to the $V$ scheme is ${Q^*}^2=1.39 \, Q^2$, 
allowing an increase in the range of pQCD.

 
 \bibitem{size}
Some results from lattice QCD, seem to indicate that 
$\alpha_s(Q=0)=0$; however, such behavior is known to be an unphysical 
artifact from finite size effects, as discussed {\it e.g.,} in Ref.~\cite{Papavassiliou:2009vk}.  An IR 
fixed point requires a vanishing derivative in the IR, $\beta(Q^2 \!= \!0) = 
0$. This likely excludes $\alpha_s(Q^2 = 0) = 0$.

 \bibitem{Papavassiliou:2009vk}
J.~Papavassiliou,
  ``Scrutinizing the Green's functions of QCD: Lattice meets Schwinger-Dyson,''
  Nucl.\ Phys.\ B, Proc. Suppl.  {\bf 199}, 44 (2010)
  [\href{http://arXiv.org/abs/0910.4487}{\tt arXiv:0910.4487} [hep-ph]].
  %%CITATION = ARXIV:0910.4487;%%
  
  \bibitem{Badalian:2000hv}
  A.~M.~Badalian,
  ``Strong coupling constant in coordinate space,''
  Phys.\ Atom.\ Nucl.\  {\bf 63}, 2173 (2000)
  [Yad.\ Fiz.\  {\bf 63}, 2269 (2000)].
  %%CITATION = YAFIA,63,2269;%%
  
   \bibitem{Peter:1997me}
  M.~Peter,
  ``The static potential in QCD: A full two-loop calculation,''
  Nucl.\ Phys.\  B {\bf 501}, 471 (1997)
  [\href{http://arXiv.org/abs/hep-ph/9702245}{\ tt arXiv:hep-ph/9702245}].
  %%CITATION = NUPHA,B501,471;%%
 An error in the expression for $a_2$  (Eq. 31) in the above reference has been corrected according to:
 %\bibitem{Schroder:1998vy}
  Y.~Schr\"oder,
  ``The static potential in {QCD} to two loops,''
  Phys.\ Lett.\  B {\bf 447}, 321 (1999)
  [\href{http://arXiv.org/abs/hep-ph/9812205}{\tt arXiv:hep-ph/9812205}];
  %%CITATION = PHLTA,B447,321;%%
 %\bibitem{Schroder:1999sg}
  Y.~Schr\"oder,
  ``The static potential in QCD,''
 Ph.~D. Thesis, Hamburg University,
1999,  Report No. DESY-THESIS-1999-021;
  %%CITATION = DESY-THESIS-1999-021;%%
%\ bibitem{Kniehl:2001ju}
  B.~A.~Kniehl, A.~A.~Penin, M.~Steinhauser and V.~A.~Smirnov,
  ``Nonabelian $\alpha(s)^3/(m(q) r^2)$ heavy-quark-antiquark potential,''
  Phys.\ Rev.\  D {\bf 65}, 091503 (2002)
  [\href{http://arXiv.org/abs/hep-ph/0106135}{\tt arXiv:hep-ph/0106135}].
  %%CITATION = PHRVA,D65,091503;%%
 
  \bibitem{Anzai:2009tm}
  C.~Anzai, Y.~Kiyo and Y.~Sumino,
  ``Static QCD potential at three-loop order,''
  Phys.\ Rev.\ Lett.\  {\bf 104}, 112003 (2010)
  [\href{http://arXiv.org/abs/0911.4335}{\tt arXiv:0911.4335} [hep-ph]].
  %%CITATION = PRLTA,104,112003;%%

\bibitem{Smirnov:2010zc}
  A.~V.~Smirnov, V.~A.~Smirnov and M.~Steinhauser,
  ``Full result for the three-loop static quark potential,''
  \href{http://arXiv.org/abs/1001.2668}{\tt arXiv:1001.2668} [hep-ph].
  %%CITATION = ARXIV:1001.2668;%%
  
  \bibitem{Brodsky:2002cx}
  S.~J.~Brodsky, D.~S.~Hwang and I.~Schmidt,
  ``Final-state interactions and single-spin asymmetries in semi-inclusive deep inelastic scattering,''
  Phys.\ Lett.\  B {\bf 530}, 99 (2002)
 [\href{http://arXiv.org/abs/hep-ph/0201296}{\tt arXiv:hep-ph/0201296}].
  %%CITATION = PHLTA,B530,99;%%

\bibitem{Collins:2002kn}
  J.~C.~Collins,
  ``Leading-twist single-transverse-spin asymmetries: Drell-Yan and Deep-Inelastic Scattering,''
  Phys.\ Lett.\  B {\bf 536}, 43 (2002)
  [\href{http://arXiv.org/abs/hep-ph/0204004}{\tt arXiv:hep-ph/0204004}].
  %%CITATION = PHLTA,B536,43;%%

\bibitem{Boer:2002ju}
  D.~Boer, S.~J.~Brodsky and D.~S.~Hwang,
  ``Initial-state interactions in the unpolarized Drell-Yan process,''
  Phys.\ Rev.\  D {\bf 67}, 054003 (2003)
 [\href{http://arXiv.org/abs/hep-ph/0211110}{\tt arXiv:hep-ph/0211110}].
  %%CITATION = PHRVA,D67,054003;%%

\bibitem{Brodsky:1995ds}
  S.~J.~Brodsky, A.~H.~Hoang, J.~H.~Kuhn and T.~Teubner,
  ``Angular distributions of massive quarks and leptons close to threshold,''
  Phys.\ Lett.\  B {\bf 359}, 355 (1995)
  [\href{http://arXiv.org/abs/hep-ph/9508274}{\tt arXiv:hep-ph/9508274}].
  %%CITATION = PHLTA,B359,355;%%
  
  %\cite{Collins:2007nk}
\bibitem{Collins:2007nk}
  J.~Collins and J.~W.~Qiu,
  ``$k_{T}$ factorization is violated in production of high-transverse-momentum particles in hadron-hadron collisions,''
  Phys.\ Rev.\  D {\bf 75}, 114014 (2007)
 [\href{http://arXiv.org/abs/0705.2141}{\tt arXiv:0705.2141} [hep-ph]].
  %%CITATION = PHRVA,D75,114014;%%
  
 \bibitem{uncertainties}
The uncertainties of the dense and sparse cross-hatched
bands are partially correlated because they share the same contribution 
from $\Lambda_{\rm QCD}$ and $\alpha_{\overline{MS}}$. The correlated contributions should be removed when checking
the consistency of the two bands. The agreement is still excellent when  only the uncertainties from the 
$\alpha_V$ series truncation at two loops for the sparse cross-hatched band and the uncertainties on the CSR series
truncation for the dense cross-hatched band are used to estimate the band widths.

\bibitem{green band}   The cross-hatched band could be computed only for  $Q^2 \gtrsim 3$ GeV$^2$ because, as already noticed, the perturbative series in the $\overline{MS}$ scheme is less suitable for low $Q^2$ studies: the leading-order scale shift in the present case is $Q^*=0.368 \, Q$. 
  

  

  
  
\end{thebibliography}
\end{document}